\RequirePackage[l2tabu,orthodox]{nag}
\expandafter\ifx\csname pdfoutput\endcsname\relax \else
  \ifnum\pdfoutput=1
  \fi
\fi
\PassOptionsToPackage
  {pdftex,pdfa,bookmarksnumbered,unicode,colorlinks=false,hidelinks}
  {hyperref}

\documentclass[conference]{IEEEtran}

\usepackage[T1]{fontenc}
\usepackage{newtxtext} 
\usepackage{amsmath}   
\usepackage{amssymb}   
\usepackage{newtxmath} 
\usepackage{microtype} 

\usepackage{balance}
\usepackage{booktabs}
\usepackage{etoolbox}
\usepackage{graphicx}
\usepackage[numbers,sort]{natbib}
\usepackage[hyphens]{url}
\usepackage{xpatch}

\usepackage[table]{xcolor}
\definecolor{gray}{rgb}{0.9, 0.9, 0.9}

\usepackage[a-2u]{pdfx}
\usepackage{hyperref}
\usepackage{cleveref}
\usepackage{bookmark}

\usepackage[utf8]{inputenc}

\newif\ifanonymized\anonymizedfalse 
\newif\ifroughdraft\roughdraftfalse  

\ifroughdraft
\newcommand{\kibitz}[3]
  {\textcolor{#1}{[\textbf{#2}\ifx&#3&\else: \textit{#3}\fi]}}
\else
\newcommand{\kibitz}[3]{}
\fi

\newcommand{\iclabcountries}{62}
\newcommand{\iclabblockedcountries}{60}
\newcommand{\iclabases}{234} 
\newcommand{\iclabvantagepoints}{281}
\newcommand{\iclabvpns}{264}
\newcommand{\iclabvpnscountries}{55}
\newcommand{\iclabvolunteer}{17}
\newcommand{\iclabvolunteercountries}{13}
\newcommand{\iclabvpsintersectioncountries}{6}
\newcommand{\iclabnotfree}{8}
\newcommand{\iclabpartlyfree}{22}
\newcommand{\iclabfree}{32}
\newcommand{\iclaburls}{45,565}
\newcommand{\iclabroundurls}{45,000}
\newcommand{\iclaboverallblocking}{3,602}
\newcommand{\iclabroundblocking}{3,500}
\newcommand{\iclabmeasurements}{53,906,532}
\newcommand{\iclabblockpagesunique}{2,782} 
\newcommand{\iclabblockpagesnum}{232,183} 
\newcommand{\iclabblockpagecountries}{50}
\newcommand{\iclabdnsnum}{15,007}
\newcommand{\iclabdnsunique}{489}
\newcommand{\iclabdnsblockedcountries}{56}
\newcommand{\iclabpacketnum}{143,225}
\newcommand{\iclabpacketunique}{1,205}
\newcommand{\iclabpacketcountries}{54}
\newcommand{\iclabrawpacketnum}{19,493,925}
\newcommand{\iclabrawpacketunique}{11,482}
\newcommand{\iclabrawpacketcountries}{55}

\newcommand{\eg}{\emph{e.g.,}}

\newcommand{\etc}{\emph{etc.}}

\newcommand{\sectionref}[1]{\textsection\ref{#1}}

\newif\iffirstuseofsystemname\firstuseofsystemnametrue
\ifanonymized
\def\sayanononce{%
  \iffirstuseofsystemname\ (anonymized)\firstuseofsystemnamefalse\fi}
\providecommand{\systemname}{[System]\sayanononce}
\providecommand{\Systemname}{[System]\sayanononce}
\else
\providecommand{\systemname}{ICLab}
\providecommand{\Systemname}{ICLab}
\fi

\providecommand{\myparab}[1]{\smallskip\noindent\textbf{#1} }

\overfullrule \ifroughdraft 5pt \else 0pt\fi

\AtBeginDocument{%
  \sfcode\csname\encodingdefault\string\textquotedblright\endcsname=0%
  \sfcode\csname\encodingdefault\string\textquoteright\endcsname=0%
}

\pdfglyphtounicode{bigcircle}{25EF}

\DeclareMicrotypeSet[protrusion]{not-tt}{
  family = {rm*, sf*}
}
\UseMicrotypeSet[protrusion]{not-tt}

\def\figurename{\textsc{Fig.}}
\def\tablename{\textsc{Table}}

\makeatletter
\let\@IEEEconsolenoticeconference\relax

\AtBeginDocument{
  \setlength{\@fptop}{0\p@}
  \setlength{\@dblfptop}{0\p@}
}
\makeatother

\setcitestyle{citesep={], [}}
\makeatletter
\xpatchcmd{\NAT@citexnum}
  {\def@NAT@last@yr{--\NAT@penalty}}   
  {\def@NAT@last@yr{]--[\NAT@penalty}} 
  {}     
  {\ddt} 
\def\NAT@def@citea{\def\@citea{\NAT@separator}}
\makeatother

\newcommand{\ICauthor}[2]{#1\forcsvlist{\IEEEauthorrefmark}{#2}}
\newcommand{\ICaffil}[5]{\parbox{#2}{\centering
    \IEEEauthorrefmark{#1}#3\\{\small \{#4\}@#5}}}
\begin{document}

\date{}

\firstuseofsystemnamefalse
\title{\Systemname: A Global, Longitudinal\\
       Internet Censorship Measurement Platform}

\ifanonymized
\author{Anonymous}
\else
\author{
  \IEEEauthorblockN{
    \ICauthor{Arian Akhavan Niaki}{1,2}\quad
    \ICauthor{Shinyoung Cho}{1,2,3}\quad
    \ICauthor{Zachary Weinberg}{1,4}\\
    \ICauthor{Nguyen Phong Hoang}{3}\quad
    \ICauthor{Abbas Razaghpanah}{3}\quad
    \ICauthor{Nicolas Christin}{4}\quad
    \ICauthor{Phillipa Gill}{2}}
  \vspace*{1ex}
  \IEEEauthorblockA{
    \ICaffil{2}{2.2in}
      {University of Massachusetts, Amherst}
      {arian, shicho, phillipa}{cs.umass.edu}
    \quad
    \ICaffil{3}{2.75in}
      {Stony Brook University}
      {shicho, nghoang, arazaghpanah}{cs.stonybrook.edu}
    \quad
    \ICaffil{4}{1.6in}
       {Carnegie Mellon University}
       {zackw, nicolasc}{cmu.edu}}
}
\fi

\maketitle

\begin{NoHyper}
\let\thefootnote\relax
\footnotetext{\IEEEauthorrefmark{1}Authors contributed equally}
\end{NoHyper}

\firstuseofsystemnametrue
\begin{abstract}
Researchers have studied Internet censorship for nearly as long as attempts to
censor contents have taken place. Most studies have however been limited to a
short period of time and/or a few countries; the few exceptions have traded
off detail for breadth of coverage. Collecting enough data for a
comprehensive, global, longitudinal perspective remains challenging.

In this work, we present \systemname, an Internet measurement platform
specialized for censorship research.  It achieves a new balance between
breadth of coverage and detail of measurements, by using commercial VPNs as
vantage points distributed around the world.  \systemname\ has been operated
continuously since late 2016.  It can currently detect DNS manipulation and
TCP packet injection, and overt ``block pages'' however they are delivered.
\systemname\ records and archives raw observations in detail, making
retrospective analysis with new techniques possible.  At every stage of
processing, \systemname\ seeks to minimize false positives and manual
validation.

Within \iclabmeasurements\ measurements of individual web pages, collected by
\systemname\ in 2017 and 2018, we observe blocking of
\iclaboverallblocking\ unique URLs in \iclabblockedcountries\ countries.
Using this data, we compare how different blocking techniques are deployed in
different regions and/or against different types of content.  Our longitudinal
monitoring pinpoints changes in censorship in India and Turkey concurrent with
political shifts, and our clustering techniques discover 48 previously unknown
block pages.  \systemname's broad and detailed measurements also expose other
forms of network interference, such as surveillance and malware injection.
\end{abstract}

\firstuseofsystemnametrue
\section{Introduction}\label{sec:introduction}

For the past 25 years, the Internet has been an important forum for people who
wish to communicate, access information, and express their opinions.  It has
also been the theater of a struggle with those who wish to control who can be
communicated with, what information can be accessed, and which opinions can be
expressed.  National governments in particular are notorious for their
attempts to impose restrictions on online
communication~\cite{oni2010controlled}.  These attempts have had unintentional
international consequences~\cite{Iljitschirootchina, anon2012cn.collateral,
  bgp2008pakistan, marczak2015cn.cannon}, and have raised questions about
export policy for network management products with legitimate uses (\eg\ virus
detection and protection of confidential information) that can also be used to
violate human rights~\cite{dalek2013.url.filtering}.

The literature is rich with studies of various aspects of Internet
censorship~\cite{Anonymous:2014, Aryan:2013, Iljitschirootchina,
Pearce:2017:Iris, VanderSloot:2018:Quack, Will:2016:Satellite,
Yadav2018:ooni-flaw, anderson2013ir.throttling, anderson2014.ripe,
anon2012cn.collateral, bgp2008pakistan, bgpmon2014turkey,
burnett2015.encore, claytonchina, dainotti2013eg.ly.outages,
dalek2013.url.filtering, ensafi2015cn.hidden, farnan2016cn.poisoning,
gill.2015.worldwide, jones.2014.blockpages, marczak2015cn.cannon,
nisar2018incentivizing, ooni-paper, parkchina, pearce.2017.augur, xuchina}
but a global, longitudinal baseline of censorship covering a variety of
censorship methods remains elusive.  We highlight three key challenges that
must be addressed to make progress in this space:

\myparab{Challenge 1: Access to Vantage Points.}  With few
exceptions,\footnote{China filters inbound as well as outbound traffic, making
external observation simpler.} measuring Internet censorship requires access
to “vantage point” hosts within the region of interest.

The simplest way to obtain vantage points is to recruit
volunteers~\cite{verkamp2012.mechanics, sfakianakis2011.censmon,
gill.2015.worldwide, ooni-paper}. Volunteers can run software that performs
arbitrary network measurements from each vantage point, but recruiting more
than a few volunteers per country and retaining them for long periods is
difficult. Further, volunteers may be exposed to personal risks for
participating in censorship research.

More recently, researchers have explored alternatives, such as employing open
DNS resolvers~\cite{Pearce:2017:Iris, Will:2016:Satellite}, echo
servers~\cite{VanderSloot:2018:Quack}, Web browsers visiting instrumented
websites~\cite{burnett2015.encore}, and TCP side
channels~\cite{pearce.2017.augur, ensafi2014detecting}.  These alternatives
reduce the risk to volunteers, and can achieve broader, longer-term coverage
than volunteer labor.  However, they cannot perform arbitrary network
measurements; for instance, open DNS resolvers can only reveal DNS-based
censorship.

\myparab{Challenge 2: Understanding What to Test.}  Testing a single blocked
URL can reveal that a censorship system exists within a country, but does not
reveal the details of the censorship policy, how aggressively it is enforced,
or all of the blocking techniques used.
Even broad test lists, like those maintained by the Citizen
Lab~\cite{testlist}, may be insufficient~\cite{darer2017filteredweb}.  Web
pages are often short-lived, so tests performed in the present may be
misleading~\cite{weinberg.2017.topics}.

\myparab{Challenge 3: Reliable Detection.}  Censors can prevent access to
content in several different ways.  For instance, censors may choose to supply
``block pages'' for some material, which explicitly notify the user of
censorship, and mimic site outages for other material (see
\sectionref{sec:overt-covert})~\cite{gill.2015.worldwide, dalek2015yemen}.

Many recent studies focus on a single technique~\cite{Pearce:2017:Iris,
VanderSloot:2018:Quack, Will:2016:Satellite, burnett2015.encore,
ensafi2014detecting, pearce.2017.augur}.  This is valuable but incomplete,
because censors may combine different techniques to filter different types of
content.

As the Internet evolves and new modes of access appear (\eg~mobile devices),
censorship evolves as well, and monitoring systems must keep
up~\cite{molavi2015identifying, anderson2013ir.throttling}.  Ad-hoc detection
strategies without rigorous evaluation are prone to false
positives~\cite{Yadav2018:ooni-flaw}. For example, detecting filtering via DNS
manipulation requires care to deal with CDNs~\cite{Will:2016:Satellite,
  Pearce:2017:Iris} and detection of block pages requires taking regional
differences in content into account~\cite{jones.2014.blockpages}.

\subsection{Contributions}

We present \systemname, a censorship measurement platform that tackles these
challenges.  \Systemname\ primarily uses commercial Virtual Private Network
servers (VPNs) as vantage points, after validating that they are in their
advertised locations.  VPNs offer long-lived, reliable vantage points in
diverse locations, but still allow detailed data collection from all levels of
the network stack. \Systemname\ also deploys volunteer-operated devices (VODs)
in a handful of locations.

\Systemname\ is extensible, allowing us to implement new experiments when new
censorship technologies emerge, update the URLs that are tested over time, and
re-analyze old data as necessary.  To date \systemname\ has only been used to
monitor censorship of the web, but it could easily be adapted to monitor other
application-layer protocols (\eg\ using techniques such as those
in~\citet{molavi2015identifying}). Besides \systemname\ itself, and its
collected data, we offer the following contributions:

\myparab{Global, longitudinal monitoring.}  Since its launch in 2016,
\systemname\ has been continuously conducting measurements in
\iclabcountries\ countries, covering \iclabases\ autonomous systems (ASes) and
testing over \iclabroundurls\ unique URLs over the course of more than two
years. The platform has detected over \iclabroundblocking\ unique URLs blocked
using a variety of censorship techniques.  We discuss our discoveries in more
detail in Section~\ref{sec:analysis:results}.

\myparab{Enhanced detection accuracy.}  \systemname\ collects data from all
levels of the network stack and detects multiple different types of network
interference.  By comparing results across all the detection techniques, we
can discover inaccuracies in each and refine them.  We have eliminated all
false positives from our block page detector.  DNS manipulation detection
achieves a false positive rate on the order of $10^{-4}$ when cross-checked
against the block page detector (see Section~\ref{sec:dnsmanipulation}).
Similar cross-checking shows a negligible false positive rate for TCP packet
injection (see Section~\ref{sec:packetinjection}).

\myparab{Semi-automated block page detection.}  We have developed a new
technique for discovering both variations on known block pages and previously
unknown block pages.  These explicit notifications of censorship are easy for
a human to identify, but machine classifiers have trouble distinguishing them
from other short HTML documents expressing an error message.  Existing systems
rely on hand-curated sets of regular expressions, which are brittle and
tedious to update.

\Systemname\ includes two novel machine classifiers for short error messages,
designed to facilitate manual review of groups of suspicious messages, rather
than directly deciding whether each is a block page.  Using these classifiers
we discovered 48 previously undetected block page signatures from 13
countries.  We describe these classifiers and their discoveries in more detail
in Section~\ref{sec:detection:clustering}.

\section{Background}\label{sec:background}

Here we briefly review the techniques used to block access to information
online, two different options for implementation, and how the censor's goals
affect their implementation choices.

\subsection{Network-level blocking techniques}\label{sec:blocking-mechanism}
Abstractly, all attempts to interfere with website access are
man-in-the-middle (MITM) attacks on communications between a web browser and
server. Depending on the location and configuration of their MITM devices,
censors may interfere with traffic outside the borders of their own
authority~\cite{bgp2008pakistan, Iljitschirootchina, farnan2016cn.poisoning}.

\myparab{DNS manipulation.} When visiting a website, the user's browser must
first resolve the web server's IP address using DNS. DNS traffic is
unencrypted, and less than 1\% of it is
authenticated~\cite{wander.2017.dnssec}. Using either DNS servers they
control, or packet injection from routers, censors can forge responses
carrying DNS error codes such as “host not found” (NXDOMAIN), non-routable IP
addresses, or the address of a server controlled by the
censor~\cite{Anonymous:2014, zittrain2003internet}.

\myparab{IP-based blocking.} Once the browser has an IP address of a web
server, it makes a TCP connection to that server. Censors can discard TCP
handshake packets destined for IP addresses known to host censored content,
reply with a TCP reset packet, or reroute them to a server controlled by the
censor~\cite{hoang:2018}.

\myparab{TCP packet injection.} Censors can also allow the TCP handshake to
complete, and then inject a packet into the TCP stream that either supersedes
the first response from the legitimate server, or breaks the connection before
the response arrives~\cite{weaver.2009.reset}.  For unencrypted websites, this
technique allows the censor to observe the first HTTP query sent by the
client, and thus block access to individual
pages~\cite{crandall2007.conceptdoppler}.

\myparab{Transparent proxy.}  Censors wishing to exercise finer control can
use a “transparent proxy” that intercepts all HTTP traffic leaving the
country, decodes it, and chooses whether or not to forward
it~\cite{dalek2013.url.filtering}. Transparent proxies act as TCP peers and
may modify HTTP traffic passing through, which makes them
detectable~\cite{Weaver:2014:HWP}.  They permit fine-grained decisions about
\emph{how} to block access to content.  However, they are specific to
unencrypted HTTP and cannot be used to censor traffic in any other protocol.

\subsection{On-path and in-path censors}

Hardware performing DNS manipulation, IP-based blocking, or TCP packet
injection can be connected to the network in two different ways.  It is not
known which option is more commonly used~\cite{verkamp2012.mechanics,
hellmeier.2016.toolkit}.

\emph{On-path} equipment observes a copy of all traffic passing through a
network link. It can react by injecting packets into the link, but cannot
modify or discard packets that are already within the flow. While on-path
techniques are relatively cheap and easy to deploy, detection is also easy, as
injected packets appear alongside legitimate traffic.

\emph{In-path} equipment operates on the \emph{actual} traffic passing through
the network link, and can inject, \emph{modify, or discard} packets.  In-path
equipment must operate at the line-rate of a backbone router, so it is more
expensive and its features may be limited (\eg\ payload inspection may not be
an option), but it is harder to detect.

\subsection{Overt and covert censorship}
\label{sec:overt-covert}

Censorship's visible effects can be either \emph{overt} or \emph{covert}. In
overt censorship, the censor sends the user a “block page” instead of the
material that was censored. In covert censorship, the censor causes a network
error that could have occurred for other reasons, and thus \emph{avoids}
informing the user that the material was censored. Censors may choose to be
overt for some material and covert for other material. For instance, Yemen has
been observed to overtly block pornography, which is illegal there, and to
covertly block disfavored, but legal, political
content~\cite{gill.2015.worldwide}.

Overt censorship can be accomplished with a transparent HTTP proxy, an
injected TCP packet or DNS response that directs the browser to a server
controlled by the censor, or by rerouting TCP traffic to a server controlled
by the censor.  Covert censorship can be accomplished with a transparent HTTP
proxy, an injected TCP reset packet, an injected DNS error or non-routable
address, or by discarding packets.

\section{System Architecture}\label{sec:iclab}

\begin{figure}
\includegraphics[width=\hsize]{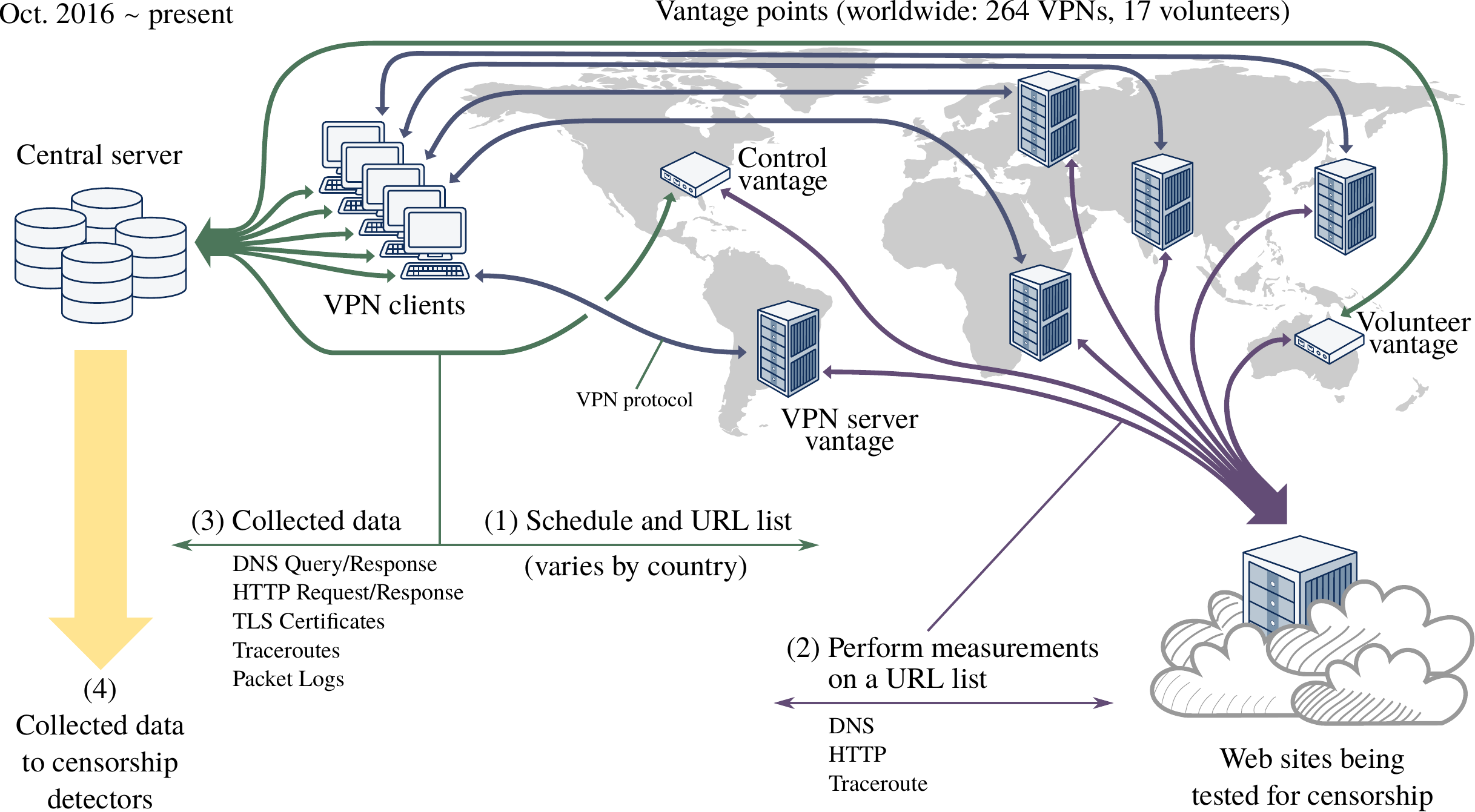}
\caption{\textsc{Architecture of \systemname.} (1)~The central server sends a
  measurement schedule along with an associated test list to vantage
  points. (2)~The vantage points perform measurements.  (3)~Collected data is
  uploaded to the central server. (4)~Censorship detection is done centrally.}
\label{fig:architecture}
\end{figure}

\Systemname\ is a platform for measuring censorship of network traffic.  As
shown in Figure~\ref{fig:architecture}, it consists of a central control
server and a set of vantage points distributed worldwide.  The central server
schedules measurements for each vantage point to perform, distributes test
lists, and collects measurement results for analysis.  The vantage points send
and receive network traffic to perform each measurement, and upload their
observations to the central server.  All analysis is done centrally after the
measurements have completed.  Raw observations, including complete packet
logs, are archived so that new analysis techniques can be applied to old data.
There are two types of vantage points: volunteer-operated devices (VODs)
configured by us and installed in locations of interest by our
volunteers,\footnote{Most of these are low-cost Raspberry~Pi devices.} and
VPN-based clients, which forward traffic through commercial VPN proxies
located in various countries.

\subsection{Design Goals}\label{sec:designgoals}

We designed \systemname\ to achieve the following properties:

\myparab{Global, continuous monitoring.}  The techniques used for Internet
censorship, the topics censored, and the thoroughness with which censorship is
enforced are known to vary both among~\cite{Pearce:2017:Iris,
dalek2013.url.filtering, darer2017filteredweb, burnett.2013.sense,
gill.2015.worldwide, hellmeier.2016.toolkit} and within
\cite{ensafi2014cn.largescale, khattak.2014.isp, aase2012whiskey,
wright2014cn.regional, xuchina} countries.  Therefore, the system should
operate vantage points in multiple locations within each of many countries, to
produce a comprehensive global view of censorship.  Censorship may ratchet
upward over time~\cite{freedom, oni2010controlled}, may change abruptly in
response to political events~\cite{dainotti2013eg.ly.outages} and may even
cease after governing parties change~\cite{gill.2015.worldwide}. Therefore,
the system should perform its measurements continuously over a period of
years, to detect these changes as they happen.

\myparab{Reproducible and extensible.}  The basic techniques for censoring
network traffic (described in \sectionref{sec:blocking-mechanism}) are
well-known~\cite{verkamp2012.mechanics, weaver.2009.reset} but new variations
appear regularly~\cite{anderson2013ir.throttling, farnan2016cn.poisoning}.
The short lifetime of “long tail” content means that the current content of a
website may bear no relationship to what it was when it was originally
censored~\cite{weinberg.2017.topics}.  Therefore, the system needs to be
extensible with new types of measurement, and should record as much
information as possible with each measurement (\eg\ packet traces and detailed
contextual information).

\myparab{Minimal risk to volunteers.}\label{sec:minimalrisk} Censorship
monitoring involves accessing material that is forbidden in a particular
country, from that country, and provoking a response from the censor. The
response we expect is one of the MITM attacks described in
\sectionref{sec:blocking-mechanism}, but legal or extralegal sanctions aimed
at the volunteer operating the vantage point are also possible.  The risk may
be especially significant for volunteers already engaged in human rights
reporting or advocacy. Use of commercial VPNs as vantage points is intended to
mitigate these risks.  VODs are only deployed in locations where we believe
legal or extralegal sanctions are unlikely, and we obtain informed consent
from the volunteers who operate them.

\subsection{Vantage Points}\label{sec:vantage.points}

Of \systemname's \iclabvantagepoints\ vantage points, \iclabvpns\ are
VPN-based, obtaining access to locations of interest via commercial VPN
services. \iclabvolunteer\ vantage points are VODs. The measurement software
is the same for both types of vantage; the only difference is that VPN-based
vantages route their traffic through a VPN while performing measurements.

\myparab{VPN-based vantages.}
\Systemname\ uses VPN-based vantages whenever possible, because of their
practical and ethical advantages.  We do not need to recruit volunteers from
all over the world, or manage physical hardware that has been distributed to
them, but we still have unrestricted access to the network, unlike, for
instance, phone or web applications~\cite{ooni-paper,burnett2015.encore}.  The
VPN operator guarantees high availability and reasonable bandwidth, and they
often offer multiple locations within a country.  For 75\% of the countries
where we use VPN-based vantages, the VPNs give us access to at least two ASes
within that country (see Appendix~\ref{sec:appendix_networks_access}).

On the ethical side, a commercial VPN operator is a company that understands
the risks of doing business in each country it operates in.  It is unlikely
that they would deploy a server in a country where the company or its
employees might suffer legal or extralegal sanctions for the actions of its
users.

A disadvantage of VPNs is that they only supply a lower bound on the
censorship experienced by individuals in each country, because their servers
are hosted in commercial data centers.  There is some evidence that network
censorship is less aggressively performed by data centers' ISPs than by
residential ISPs~\cite{aceto2016it.3g4g, xuchina}.  According to the CAIDA AS
classification~\cite{CAIDA-ASClass}, 41\% of the networks hosting our
VPN-based vantages are “content” networks, which are the most likely to be
subject to reduced levels of censorship.  However, we have visibility into at
least one other type of AS in 83\% of the countries we can observe.  In
countries where we have both VPNs and VODs, we have observed identical block
pages from both, indicating that all types of ASes are subject to similar
blocking policies in those countries.

User-hosted VPNs (\eg\ Geosurf~\cite{geosurf}, Hola~\cite{hola},
Luminati~\cite{luminati}) would offer access to residential ISPs, but
\systemname\ does not use them, as they have all the ethical concerns
associated with VODs, with less transparency.  Also, there are reports of
illicit actions by the operators of these VPNs, such as deploying their
software as a viral payload, and facilitating distributed denial of service
(DDoS) attacks~\cite{mi2019resident}, making it even more unethical to use
these services.

Since commercial VPNs' advertised server locations cannot be relied
on~\cite{Weinberg:2018:CPL:Proxy}, we validate their locations using
round-trip time measurements (see Appendix~\ref{sec:appendix_geoloc} for
details), and we only use the servers whose locations are accurately
advertised.

\myparab{Volunteer-operated device vantages.}  VODs are more difficult to keep
running, and require a local volunteer comfortable with the risks associated
with operating the device.  Since \systemname\ does not collect personally
identifiable information \emph{about} the volunteers, our IRB has determined
that this project is not human subjects research.  However, we are guided by
the principles of ethical human subjects research, particularly the need to
balance potential benefits of the research against risks undertaken by
volunteers. Most of our VODs have been deployed opportunistically through
collaborations with NGOs and organizations interested in measuring Internet
censorship from a policy perspective.  For each deployed VOD, we maintain
contact with the volunteer, and monitor the political situation in the country
of deployment.  We have deemed some countries too risky (for now) to recruit
volunteers in (\eg~Iran, Syria).

\begin{table}
  \caption{Country Coverage of \systemname. \ulcshape The number of countries
    and ASes on each continent where we have vantage points with validated
    locations, since 2017.  Oceania includes Australia.  VPNs: virtual private
    network servers. VODs: volunteer-operated devices. NF,~PF,~F: of the
    countries with vantage points, how many are politically not~free,
    partially~free, or free (see Appendix~\ref{sec:appendix_fscore}).}
  \label{tab:iclab_vps}
  \centering
    \footnotesize
    \begin{tabular}{lrrrrrrr}
    \toprule
    \scriptsize\bfseries Continent &
    \scriptsize\bfseries VPNs &
    \scriptsize\bfseries VODs &
    \scriptsize\bfseries Countries &
    \scriptsize\bfseries ASes &
    \scriptsize\bfseries NF &
    \scriptsize\bfseries PF &
    \scriptsize\bfseries F \\
    \midrule
    Asia       &  64 &  4 & 14/32 &  54 & 5 &  7 &  2 \\
    Africa     &   9 & 10 &  9/72 &  19 & 1 &  6 &  2 \\
    N. America &  87 &  1 &  5/17 &  81 & 0 &  1 &  4 \\
    S. America &   9 &  0 &  5/20 &   6 & 1 &  3 &  1 \\
    Europe     &  83 &  2 & 27/42 &  64 & 1 &  5 & 21 \\
    Oceania    &  12 &  0 &  2/\phantom{0}6 &  11 & 0 &  0 &  2 \\
    \addlinespace
    Total& \iclabvpns & \iclabvolunteer&  \iclabcountries/189 & \iclabases & \iclabnotfree
                                & \iclabpartlyfree & \iclabfree \\
    \bottomrule
    \end{tabular}
\end{table}

\myparab{Breadth of coverage.}
As of this writing, \systemname\ has VPN-based vantage points in
\iclabvpnscountries\ countries, and volunteer-operated clients in
\iclabvolunteercountries\ countries.
\iclabvpsintersectioncountries\ countries host both types of clients, so
\systemname\ has vantage points in \iclabcountries\ countries overall.
\Systemname\ seeks to achieve both geographic and political diversity in its
coverage.  Table~\ref{tab:iclab_vps} summarizes our
current geographic diversity by continent, and political diversity by a
combination of two scores of political freedom, developed by Freedom
House~\cite{freedom} and Reporters Without Borders~\cite{rwb} (see
Appendix~\ref{sec:appendix_fscore}).

It is easier to acquire access to vantage points in Europe, North America, and
East Asia than in many other parts of the world.  We have plans for expanded
coverage in Africa and South America in the near future, via additional VPN
services.  It is also easier to acquire access to vantage points in “free” and
“partially free” than “not free” countries, because it is often too risky for
either VPN services or volunteer-operated devices to operate in “not free”
countries.  Expanding our coverage of “not free” countries is a priority for
future development of \systemname, provided we can do it safely.

Internet censorship does happen in the “partly free” and “free” countries, and
is not nearly as well documented as it is for specific “not free” countries
(most notably China).  Our broad coverage of these classes of countries gives
us the ability to track changes over time, across the full spectrum of
censorship policy, worldwide.

\subsection{Test Lists}\label{s:test_lists}

At present, \systemname's measurements are focused on network-level
interference with access to websites.  \Systemname's vantage points test
connectivity to the websites on three lists: the Alexa global top 500 websites
(ATL)~\cite{alexa.undated.topsites}, the websites identified as globally
sensitive by the Citizen Lab~\cite{testlist} and the Berkman Klein
Center~\cite{berkman-testlist}\footnote{The lists maintained by Citizen Lab
  and Berkman Klein are formally independent but have substantial overlap, so
  we combine them.} (\mbox{CLBL-G}), and, for each country, the websites identified
as locally sensitive in that country by Citizen Lab and Berkman Klein
(\mbox{CLBL-C}).  We only use the global top 500 sites from Alexa's ranking, because
its “long tail” is unstable~\cite{Scheitle:2018:TopList, LePochat2019}.  All
test lists are updated weekly.

\Systemname\ has tested a total of 47,000 unique URLs over the course of its
operation.  Because all of the vantage points test ATL and \mbox{CLBL-G}, there is
more aggregated data for these sites: 40\% of our data is from sites on ATL,
40\% from sites on \mbox{CLBL-G}, and 20\% from sites on \mbox{CLBL-C}.  Individual vantage
points test anywhere from 3,000 to 5,700 URLs per measurement cycle, depending
on the size of \mbox{CLBL-C} for the vantage point's country.  This is by no means
the complete set of sites blocked in any one
country~\cite{darer2017filteredweb}, and we have plans to broaden our testing,
as described further in Section~\ref{sec:discussion}.

\subsection{Data Collection}
\label{subsec:data_collection}

A \emph{measurement} of a URL is an attempt to perform an HTTP GET
request to that URL, recording information about the results from
multiple layers of the network stack: (1) The complete DNS request and
response or responses for the server hostname (using both a local
resolver and a public DNS resolver). (2) Whether or not a TCP
connection succeeded. (3) For HTTPS URLs, the certificate chain
transmitted by the server. (4) The full HTTP response (both headers
and body). (5) A traceroute to the server. (6) A comprehensive packet
trace for the duration of the measurement. This allows us to identify
anomalies that would not be apparent from application-layer
information alone. For instance, when packets are injected by on-path
censors, we can observe both the injected packets and the legitimate
responses they conflict with (see Section~\ref{sec:packetinjection}).

Each vantage point measures connectivity to all of the sites on its test list
at least once every three days, on a schedule controlled by the central
server.  Depending on the size of the test list, a cycle of measurements
typically runs for 1--2 hours.

\begin{figure}
\includegraphics[page=1,width=0.98\hsize]{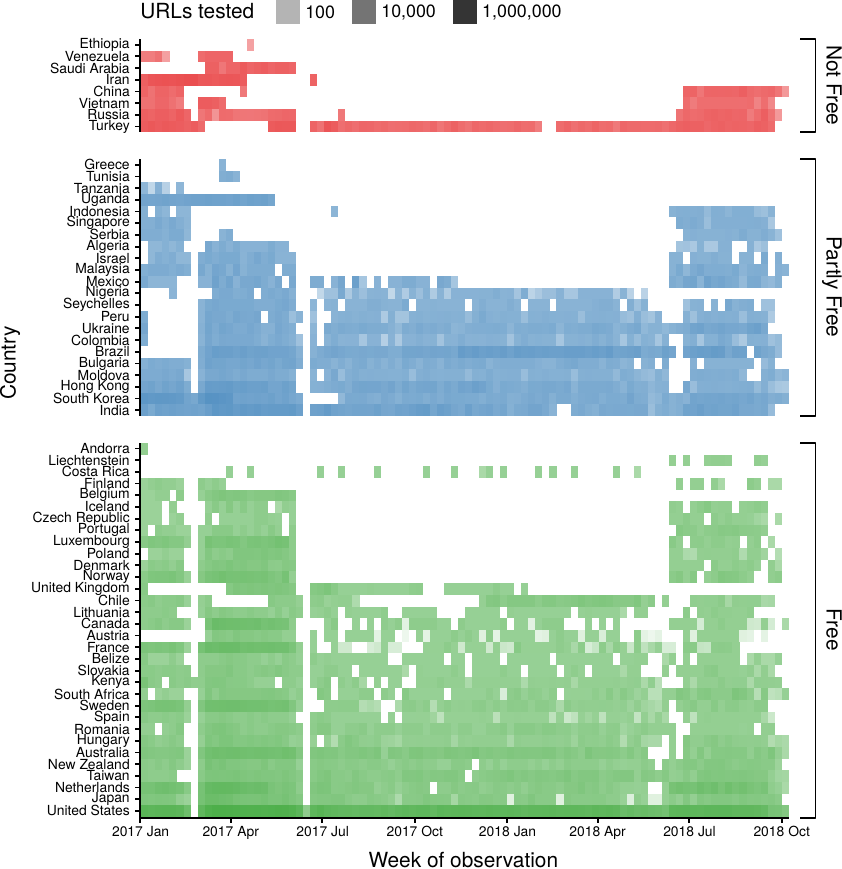}
\caption{\textsc{Measurements since 2017 by country.}  For each of the
    \iclabcountries\ countries where we have, or had, vantage points since
    2017, the total number of measurements per week.}
\label{fig:scatter}
\end{figure}

Figure~\ref{fig:scatter} depicts \systemname's measurements over time in each
country.  Operating \systemname\ over a multi-year period has not been easy;
several outages are visible in Figure~\ref{fig:scatter}.  For instance, we
lost access to our vantage points in Iran in May 2017 due to a change in the
international sanctions imposed on Iran, and we suffered a year-long,
multi-country outage due to one VPN provider making configuration changes
without notice.  The latter incident led us to improve our internal monitoring
and our tracking of VPN configuration changes.

Between January 2017 and September 2018, \systemname\ conducted
\iclabmeasurements\ measurements of \iclaburls\ URLs in
\iclabcountries\ countries and \iclabases\ ASes.
We publish our data for use by other researchers,\footnote{%
\ifanonymized URL omitted for anonymity.%
\else Available online at \url{https://iclab.org/}.
\fi} with periodic updates as we continue operation.

\subsection{Control Nodes} \label{subsec:control_nodes}

Many tests of censorship rely on comparison of measurements between the
vantage point and a ``control'' location, where there is not anticipated to be
censorship.  We repeat all the measurements performed by our vantage points on
a \emph{control node} located in an academic network in the USA.  This network
allows access to all the sites we test for accessibility.  The control node
has also suffered outages.  In this paper, we use public data sets compiled by
other researchers to fill in the gaps, as described in
Section~\ref{sec:detection}.  We have since deployed three more control nodes
in Europe, Asia and the USA to improve reliability and geographic diversity.

\section{Censorship Detection}
\label{sec:detection}

Next, we describe how \systemname\ detects manipulated DNS
responses~(\sectionref{sec:dnsmanipulation}), packets injected into
TCP streams~(\sectionref{sec:packetinjection}) and HTML-based block
pages~(\sectionref{sec:blockpages}).  All of \systemname's detection
algorithms are designed to minimize both false negatives, in which a
censored site is not detected, and false positives, in which ordinary
site or network outages, or DNS load balancing are misidentified as
censorship~\cite{weaver.2009.reset, jones.2014.blockpages,
ensafi2014detecting, gill.2015.worldwide}.

\subsection{DNS Manipulation}
\label{sec:dnsmanipulation}

To access a website, the browser first resolves its IP address with a
DNS query.  To detect DNS manipulation, \systemname\ records the DNS
responses for each measurement, and compares them with responses to
matching DNS queries from our control node, and with DNS responses
observed by control nodes OONI~\cite{ooni-paper} operates.
\Systemname\ applies the following heuristics, in order, to the
observations from the vantage point and the control nodes.

\myparab{Vantage point receives two responses with different ASes.}
If a vantage point receives two responses to a DNS query, both with
globally routable addresses, but belonging to two different ASes, we
label the measurement as DNS manipulation.  This heuristic detects
on-path censors who inject a packet carrying false
addresses~\cite{Anonymous:2014}.  Requiring the ASes to differ
avoids false positives caused by a DNS load balancer picking a
different address from its pool upon retransmission.

\myparab{Vantage point receives NXDOMAIN or non-routable address.}
If a vantage point receives either a ``no such host'' response to a
query (NXDOMAIN, in DNS protocol terms~\cite{rfc8020}), or an address
that is not globally routable (\eg\ \texttt{10.x.y.z})~\cite{rfc8190},
but the control nodes consistently receive a globally routable address
(not necessarily the same one) for the domain name, over a
period of seven days centered on the day of the vantage point's
observation, we label the test as DNS manipulation.  The
requirement for consistency over seven days is to avoid false
positives on sites that have been shut down, during the period where a
stale address may still exist in DNS caches.

\myparab{Vantage point receives addresses from the same AS as control nodes.}
If a vantage point receives a globally routable address, and the
control nodes also receive globally routable addresses assigned to the
same AS (not necessarily the exact same address), we label the
measurement as \emph{not} DNS manipulation.  Variation within a single
AS is likely to be due to load-balancing over a server pool in a
single location.

\myparab{Vantage point and control nodes receive addresses in
  different ASes.}  The most difficult case to classify is when the
vantage point and the control nodes receive globally routable
addresses assigned to different ASes.  This can happen when DNS
manipulation is used to redirect traffic to a specific server (\eg\ to
display a block page).  However, it can also happen when a content
provider or CDN directs traffic to data centers near the
client~\cite{Pearce:2017:Iris}.

\begin{figure}
\includegraphics[width=\hsize] {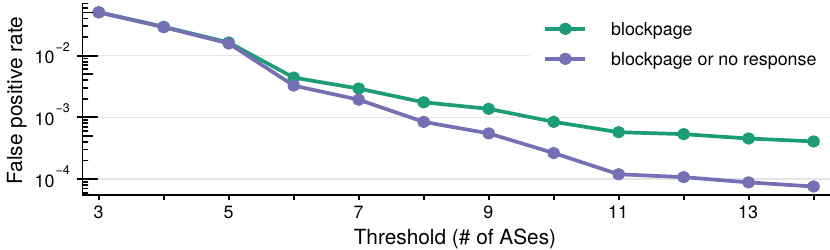}
\caption{\textsc{DNS manipulation false positives}.  The false
  positive rate for the DNS manipulation detector, as a function of
  the threshold parameter $\theta$.}
\label{fig:fpr}
\end{figure}

We distinguish censors from CDNs using the observation that censors
tend to map many blocked websites onto just a few
addresses~\cite{anon2012cn.collateral, gill.2015.worldwide}.  If a set
of websites resolve to a single IP address from the vantage point, but
resolve to IPs in more than~$\theta$ ASes from the control nodes, we
count those websites as experiencing DNS manipulation. $\theta$ is a
tunable parameter which we choose by cross-checking whether these
measurements also observed either a block page or no HTTP response at
all.  Taking this cross-check as ground truth, Figure~\ref{fig:fpr}
shows how the false positive rate for DNS manipulation varies with
$\theta$.  For the results in Section~\ref{sec:analysis:results}, we
use a conservative $\theta = 11$ which gives a false positive rate on
the order of $10^{-4}$.

\subsection{TCP Packet Injection}
\label{sec:packetinjection}

Censors may also allow DNS lookup to complete normally, but then
inject packets that disrupt the TCP handshake or subsequent traffic.
\Systemname\ detects this form of censorship by recording packet
traces of all TCP connections during each test, and analyzing them for
(1) evidence of packet injection, and (2) evidence of intent to censor
(\eg\ block page content or TCP reset flags in injected packets).  By
requiring both types of evidence, we minimize false positives.  Short
error messages delivered by the legitimate server will not appear to
be injected, and packets that, for innocuous reasons, appear to be
injected, will not display an intent to censor.

\myparab{Evidence of packet injection.}
If an end host receives two TCP packets with valid checksums and the
same sequence number but different payloads, the operating system will
generally accept the first packet to arrive, and discard the
second~\cite{rfc:tcp}.  An on-path censor can therefore suppress the
server's HTTP response by injecting a packet carrying its own HTTP
response (or simply an RST or FIN), timed to arrive first.  Because
\systemname\ records packet traces, it records both packets and
detects a conflict.  This is not infallible proof of packet injection;
it can also occur for innocuous reasons, such as HTTP load
balancers that do not send exactly the same packet when they
retransmit.

\myparab{Intent to censor: RST, FIN, or block page.}
When we detect a pair of conflicting packets, we inspect them for
evidence of intent to censor.  An injected packet can disrupt/censor
communication by carrying a TCP reset (RST) or
close (FIN) flag, causing the client to abort the connection and
report a generic error~\cite{weaver.2009.reset,
  crandall2007.conceptdoppler};  or it can carry an HTTP response
declaring the site to be censored (a “block page,” discussed further
in Section~\ref{sec:blockpages}), which will be rendered instead of
the true contents of the page the client
requested~\cite{dalek2013.url.filtering, jones.2014.blockpages}.

\begin{figure}
\centering\includegraphics[width=\hsize]{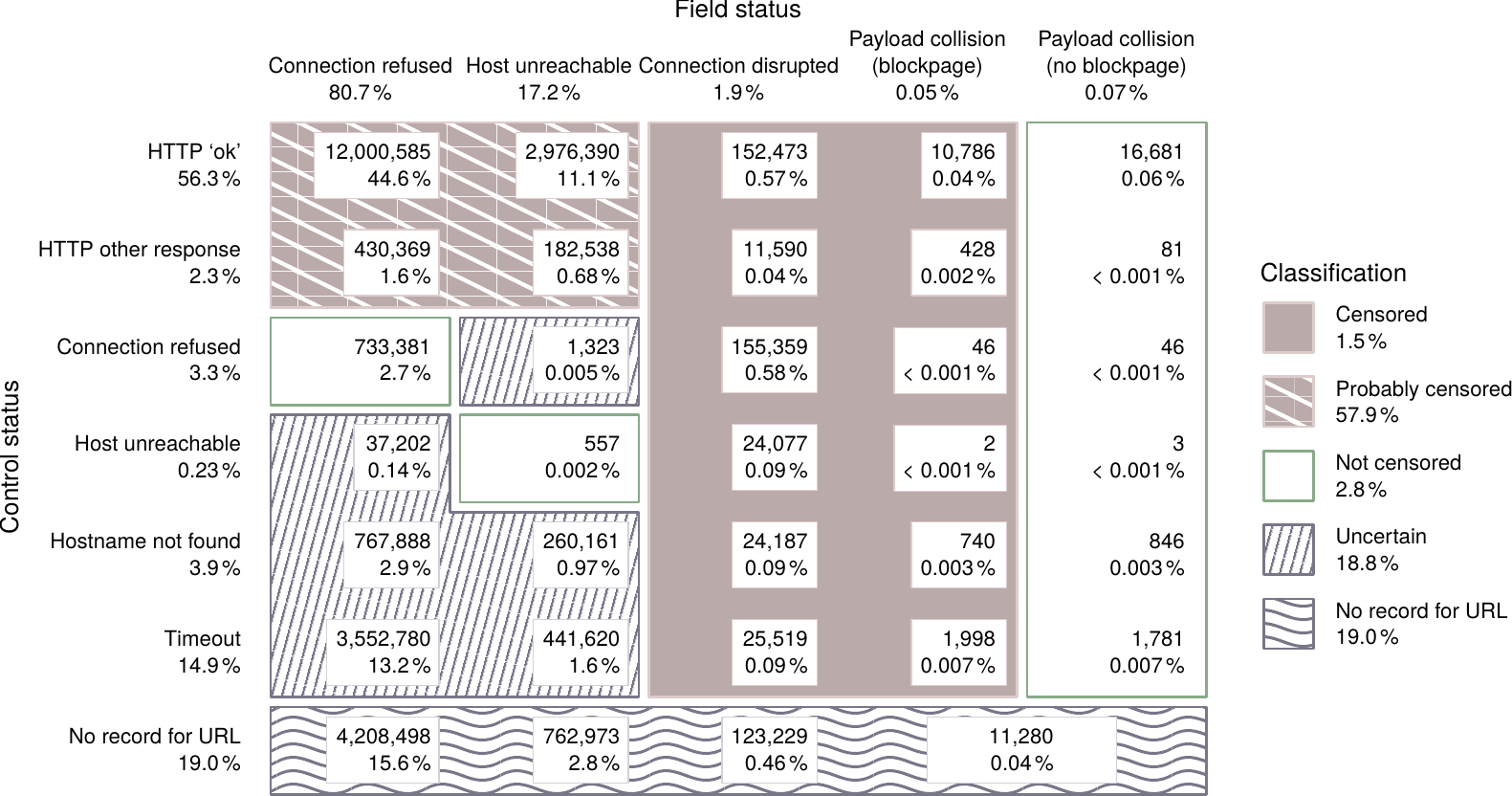}
\caption{Classification of packet anomalies by comparison to control
  observations.}
\label{f:packet-anomaly-matrix}
\end{figure}

As with DNS manipulation, we compare each observation from a vantage
point that shows evidence of packet injection, with matching
observations from a control node.  We apply the following heuristics,
in order, to pairs of observations.  The various outcomes of these
heuristics are shown in Figure~\ref{f:packet-anomaly-matrix}.

\myparab{No matching control observation.}
When a TCP stream from the vantage point shows evidence of packet
injection, but does not seem to correspond to any observation taken by
the control node, we abandon any attempt to classify it.  This is the
“No record for URL” row of Figure~\ref{f:packet-anomaly-matrix}.

This filtering is necessary because of a limitation in our
packet trace analyzer.  When a website transfers all of its
traffic to another domain name, either via a CNAME record in DNS
or using HTTP redirects, the trace analyzer cannot tell that TCP
connections to the second domain name are associated with an attempt
to test the first domain name. We conservatively do not consider
these cases as censorship.


\myparab{Packet collision after handshake, with RST or FIN.}
When a TCP stream from the vantage point shows evidence of 
collisions in TCP sequence numbers after successful 
completion of the three-way handshake, one
side of the collision has its RST or FIN bit set and the other side
has neither bit set, we label the measurement as censored by packet
injection, regardless of what the control node observed.  This is the
“connection disrupted” column of Figure~\ref{f:packet-anomaly-matrix}.
We have high confidence that all of these are
true positives.

\myparab{Packet collision after handshake, with payload conflict.}
When a TCP stream from the vantage point shows evidence of TCP
sequence number collisions after successful completion of the
three-way handshake, but neither side of the collision has the RST or
FIN bit set, we inspect the contents of the packets for a block page
signature, as described in \sectionref{sec:blockpages}.  We label the
measurement as censored by packet injection only if a known block page
signature appears in one of the packets.  These cases are the “payload
collision (blockpage)” and “payload collision (no~blockpage)” columns
of Figure~\ref{f:packet-anomaly-matrix}.  Again, we have high
confidence that these are true positives and negatives.

\myparab{Matching RST or ICMP unreachable instead of SYN-ACK.}
When a TCP SYN from the vantage point receives either a TCP RST or an
ICMP unreachable packet, instead of a SYN-ACK, and the control node
observes the same network error, we conclude the site is down
for everyone, and label the measurement as \emph{not} censored.  These cases
are the matching “connection refused” and “host unreachable” cells on
the left-hand side of Figure~\ref{f:packet-anomaly-matrix}, and we
have high confidence that they are true negatives.

\myparab{RST or ICMP unreachable instead of SYN-ACK, at vantage only.}
When a TCP SYN from the vantage point receives either a TCP RST or an
ICMP unreachable packet in response, instead of a SYN-ACK, but the
control node is able to carry out a successful HTTP dialogue, this
\emph{probably} indicates IP-based censorship observed by the vantage
point.  However, there are other possible explanations, such as a
local network outage at the vantage point, or a site blocking access
from specific IP addresses on suspicion of
malice~\cite{McDonald:2018:403}.  Manual spot-checking suggests that
many, but not all, of these observations are censorship.  These cases
are labeled as “probable censorship” in
Figure~\ref{f:packet-anomaly-matrix}, and we discuss them separately
in Section~\ref{sec:analysis:results}.

\myparab{Mismatched network errors, or timeout or DNS error at control
  node.}
When the vantage point and the control node both received a network
error in response to their initial SYN, but not the same network
error; when the control node's initial SYN received no response at
all; and when the control node was unable to send a SYN in the first
place because of a DNS error; we cannot say whether the measurement
indicates censorship.  These cases are the cells labeled “uncertain”
in the lower left-hand corner of Figure~\ref{f:packet-anomaly-matrix}.
We are conservative and do not consider these as censorship in our analysis.

\subsection{Block Page Detection and Discovery}
\label{sec:blockpages}
\label{sec:detection:lengthdiff}
\label{sec:detection:tagfreq}
\label{sec:detection:clustering}

Block page contents vary depending on the country and the technology
used for censorship. Known block pages can be detected with regular
expressions applied to the TCP payloads of suspicious packets, but
these will miss small variations from the expected text, and are no
help at all with unknown block pages.

Nonetheless, \systemname\ uses a set of 308 regular expressions to
detect known block pages.  We manually verified these match specific,
known block pages and nothing else.  40 of them were developed by the
Citizen Lab~\cite{citizenlab-blockpage}, 24 by OONI~\cite{ooni-paper},
144 by Quack~\cite{VanderSloot:2018:Quack}, and 100 by us.

Anomalous packets that do \emph{not} match any of these regular
expressions are examined for block page variations and unknown block
pages, as described below; when we discover a block page that was
missed by the regular expressions, we write new ones to cover them.

\myparab{Self-contained HTTP response.} To deliver a block page, the
protocol structure of HTTP requires a censor to inject a single packet
containing a complete, self-contained HTTP response.  This packet must
conflict with the first data packet of the legitimate response.
Therefore, only packets which are both involved in a TCP sequence
number conflict, and contain a complete HTTP response, are taken as
candidate block pages for the clustering processes described next.

\myparab{HTML structure clusters.}  The HTML tag structure of a block
page is characteristic of the filtering hardware and software used by
the censor. When the same equipment is used in many different
locations, the tag structure is often an exact match, even when the
text varies.  We reduce each candidate block page to a vector of HTML
tag frequencies (1 \verb|<body>|, 2 \verb|<p>|, 3 \verb|<em>|,
\etc)\ and compare the vectors to all other candidate block pages'
vectors, and to vectors for pages that match the known block page
regular expressions.  When we find an exact match, we manually inspect
the matching candidates and decide whether to add a new regular
expression to the detection set. Using this technique we discovered 15
new block page signatures in five countries.

\myparab{Textual similarity clusters.}  Within one country, the legal
jargon used to justify censoring may vary, but is likely to be similar
overall.  For example, one Indian ISP refers to “a court of competent
jurisdiction” in its block pages, and another uses the phrase “Hon'ble
Court” instead.  Small variations like this are evidently the same
page to a human, but a regular expression will miss them.  We apply
\emph{locality-sensitive hashing} (LSH)~\cite{zhu.2016.lsh-ensemble}
to the text of the candidate block pages, after canonicalizing the
HTML structure.  LSH produces clusters of candidate pages, centered on
pages that do match the known block page regular expressions.  As with
the tag frequency vectors, we manually inspect the clusters and decide
whether to add new regular expressions to the curated set.  Using this
technique, we discovered 33 new block page signatures in eight
countries.  An example cluster is shown in
Appendix~\ref{sec:detailed-results}.

\myparab{URL-to-country ratio.}  To discover wholly unknown block
pages we take each LSH cluster that is \emph{not} centered on a known
block page, count the number of URLs that produced a page in that
cluster, and divide by the number of countries where a page in that
cluster was observed.  This is essentially the same logic as counting
the number of websites that resolve to a single IP from a test vantage
point but not a control vantage point, but we do not use a threshold.
Instead, we sort the clusters from largest to smallest URL-to-country
ratio and then inspect the entire list manually.  The largest ratio
associated with a newly discovered block page was 286 and the smallest
ratio was 1.0.

\section{Findings}
\label{sec:analysis:results}

Between January 2017 and September 2018, \systemname\ conducted
\iclabmeasurements\ measurements of \iclaburls\ URLs in
\iclabcountries\ countries.  Because we do not have continuous coverage of all
these countries (see \sectionref{subsec:data_collection}), in this paper we
present findings only for countries where we successfully collected at
least three months' worth of data prior to September 2018.  Among those
countries, five stand out as conducting the most censorship overall: Iran,
South Korea, Saudi Arabia, India, and Kenya.  When considering specific
subsets of our data, sometimes Turkey or Russia displaces one of these five.

\subsection{Specific Results}\label{sec:specific}

We first present details of our observations for each of the three censorship
techniques that we can detect.

\myparab{DNS manipulation.} We observe \iclabdnsnum\ DNS
manipulations in \iclabdnsblockedcountries\ countries, applied to
\iclabdnsunique\ unique URLs.  98\% of these cases received NXDOMAIN
or non-routable addresses.

\begin{figure}
\centering \includegraphics {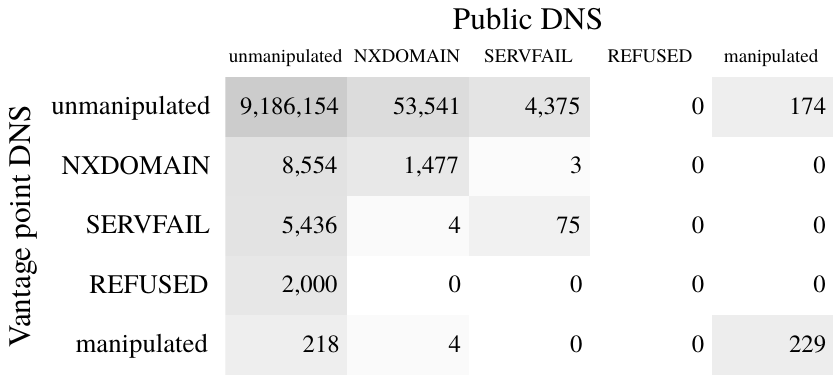}
\caption{\textsc{Comparison of DNS responses for the same domain between local and public
nameservers from the same vantage point.}}
\label{fig:dns_public_local}
\end{figure}

Figure~\ref{fig:dns_public_local} compares DNS responses from a
vantage point’s local recursive resolver with those received by the
same vantage point from a public DNS utility (\eg~\texttt{8.8.8.8}).
The upper left-hand cell of this chart counts cases where there is no
DNS censorship; the other cells in the left-hand column count cases
where censorship is being performed by the local DNS recursive
resolver.  The top rightmost cell counts the number of observations
where censorship is being performed only by a public DNS utility, and
the bottom rightmost cell counts cases where censorship is being
performed by both a local recursive resolver and a public DNS utility.
We observe censorship by public DNS utilities only for a few sites
from Russia, Bulgaria, and Iran.  The middle three columns could be
explained as either censorship or as unrelated DNS failures.

\myparab{Packet injection.}  We observe \iclabrawpacketnum\ TCP packet
injections across \iclabrawpacketcountries\ countries, applied to
\iclabrawpacketunique\ unique URLs.  However, after applying the
filtering heuristics described in \sectionref{sec:packetinjection},
only 0.7\% of these are definitely due to censorship:
\iclabpacketnum\ injections, in \iclabpacketcountries\ countries,
applied to \iclabpacketunique\ unique URLs.  (The numbers in
Figure~\ref{f:packet-anomaly-matrix} are higher because they do not
account for all the filtering heuristics.) Packet injections are
usually used to disrupt a connection without delivering a block page;
block pages are delivered by only 3.4\% of the injections we attribute
to censorship.

Another 15,589,882 packet injections---58\% of the total---are network
errors received instead of a SYN-ACK packet.  These are described as
“probable censorship” in Figure~\ref{f:packet-anomaly-matrix}.  They
could indicate an in-path censor blocking hosts by IP address, but
there are many other possible explanations.  Our synthetic results
(below) might be quite different if we were able to classify these
more accurately.

\myparab{Block pages.}  We observe \iclabblockpagesnum\ block pages across \iclabblockpagecountries\
countries, applied to \iclabblockpagesunique\ unique URLs.  Iran presents block
pages for 24.9\% of the URLs it censors, more than any other country.  In all
of the countries we monitor, block pages are most likely to be used for URLs
in the pornography and news categories (see below).

\begin{table*}
\caption{Censorship by Test List and Category.
\ulcshape For each of the three test lists we use
(see \sectionref{s:test_lists}),
the five countries censoring the most URLs from that list,
the top three FortiGuard categories for their censored URLs
(abbreviations defined in Table~\ref{tab:fortiguard_categories}),
and the percentage of URLs from that list that are censored.}
\label{tab:category_table}
\label{category_testlist}
\centering\footnotesize
{\def\nsep{\hskip 0.75\tabcolsep}
\begin{tabular}{ll@{\nsep}rll@{\nsep}rll@{\nsep}rll@{\nsep}r}
\toprule
\multicolumn{3}{c}{\textbf{Overall}} &
\multicolumn{3}{c}{\textbf{Alexa Global (ATL)}} &
\multicolumn{3}{c}{\textbf{Globally Sensitive (\mbox{CLBL-G})}} &
\multicolumn{3}{c}{\textbf{Per-Country Sensitive (\mbox{CLBL-C})}} \\
\textbf{Country} & \textbf{Category} & \textbf{Pct.} &
\textbf{Country} & \textbf{Category} & \textbf{Pct.} &
\textbf{Country} & \textbf{Category} & \textbf{Pct.} &
\textbf{Country} & \textbf{Category} & \textbf{Pct.} \\
\cmidrule(r){1-3}\cmidrule(r){4-6}\cmidrule(r){7-9}\cmidrule{10-12}

Iran            & NEWS & 13.1\% & Iran         & NEWS & 14.0\%  & Iran        & PORN  & 11.6\%  & Iran        & NEWS  & 21.0\% \\
                & PORN &  9.2\% &              & PORN & 12.7\%  &             & NEWS  &  9.4\%  &             & BLOG  & 17.6\% \\
                & BLOG &  7.5\% &              & ENT  & 10.3\%  &             & PROX  &  6.8\%  &             & POL   &  7.2\% \\
\cmidrule(r){1-3}\cmidrule(r){4-6}\cmidrule(r){7-9}\cmidrule{10-12}
South Korea     & PORN & 15.4\% & South Korea  & SHOP & 14.2\%  & Saudi Arabia& PORN & 31.0\%   & India       & ENT   & 19.0\% \\
                & NEWS &  8.4\% &              & PORN & 13.7\%  &             & GAMB & 13.5\%   &             & STRM  & 14.3\% \\
                & ORG  &  7.4\% &              & NEWS & 10.8\%  &             & PROX & 12.2\%   &             & NEWS  & 10.8\% \\
\cmidrule(r){1-3}\cmidrule(r){4-6}\cmidrule(r){7-9}\cmidrule{10-12}
Saudi Arabia    & PORN & 29.5\% & Saudi Arabia & PORN & 70.0\%  & South Korea & PORN  & 15.6\%  & Saudi Arabia& NEWS  & 54.0\% \\
                & NEWS & 11.3\% &              & ILL  & 6.6\%   &             & ORG   & 10.4\%  &             & POL   &  7.7\% \\
                & GAMB & 10.1\% &              & GAMB & 6.6\%   &             & NEWS  &  5.7\%  &             & RELI  &  7.7\% \\
\cmidrule(r){1-3}\cmidrule(r){4-6}\cmidrule(r){7-9}\cmidrule{10-12}
India           & ENT  & 13.3\% & Turkey       & PORN & 66.0\%  & Kenya       & PORN  & 14.5\% & Russia       & BLOG  & 16.5\% \\
                & STRM & 10.8\% &              & ILL  &  4.0\%  &             & GAMB  & 10.8\% &              & NEWS  & 14.4\% \\
                & NEWS & 10.4\% &              & FILE &  4.0\%  &             & PROX  &  9.0\% &              & GAMB  & 12.4\% \\
\cmidrule(r){1-3}\cmidrule(r){4-6}\cmidrule(r){7-9}\cmidrule{10-12}
Kenya           & PORN & 15.5\% & India        & ILL  & 35.5\%  & Turkey      & PORN   & 47.0\% & Turkey      & NEWS  & 29.4\% \\
                & GAMB & 10.1\% &              & IT   &  8.8\%  &             & GAMB   & 22.6\% &             & PORN  & 13.7\% \\
                & PROX &  8.3\% &              & STRM &  6.6\%  &             & ILL    &  3.2\% &             & GAMB  &  9.8\% \\
\bottomrule
\end{tabular}}
\end{table*}


\begin{table}
\caption{Censorship Variation by Technique.
\ulcshape
For each of the three techniques we can detect, the five countries
observed to censor the most URLs using that technique, and the top three
FortiGuard categories for those URLs (abbreviations defined
in Table~\ref{tab:fortiguard_categories}).  Percentages are of all unique URLs
tested.}
\label{tab:technique_country}
\centering
\begin{tabular}{lllr}
\toprule
\textbf{Technique} &
\textbf{Country} &
\textbf{Categories} &
\textbf{Pct.} \\
\midrule
Block page           & Iran         & NEWS, PORN, BLOG & 24.95\% \\
                     & Saudi Arabia & PORN, NEWS, GAMB & 11.1\% \\
                     & India        & ENT,  STRM, NEWS  &  6.4\% \\
                     & Kenya        & PORN, GAMB, PROX &  4.8\% \\
                     & Turkey       & PORN, GAMB, NEWS &  4.6\% \\
\midrule
DNS                  & Iran         & BLOG, PORN, PROX &  5.5\% \\
manipulation         & Uganda       & PORN, ADUL, LING &  1.7\% \\
                     & Turkey       & ILL, GAMB, STRM  &  0.3\% \\
                     & Bulgaria     & ILL, ARM, DOM  &  0.2\% \\
                     & Netherlands  & ILL, IM, DOM &  0.2\% \\

\midrule
TCP packet           & South Korea  & PORN, ORG, NEWS &  9.3\% \\
injection            & India        & NEWS, ILL, IT  &  2.3\% \\
                     & Netherlands  & NEWS, SEAR, GAME  &  0.9\% \\
                     & Japan        & NEWS, GAME, SEAR  &  0.9\% \\
                     & Australia    & SEAR, NEWS,  ILL  &  0.8\% \\
\bottomrule
\end{tabular}
\end{table}

\begin{table}
\caption{FortiGuard Categories and Abbreviations.
  \ulcshape
  The 25 most common categories for the URLs
  on our test lists that were censored at least once, with
  the abbreviated names used in Tables~\ref{tab:category_table}
  and~\ref{tab:technique_country}, and the percentage of URLs
  in each category.  CLBL includes both global and per-country test lists.}
\label{tab:fortiguard_categories}
\centering\footnotesize
\begin{tabular}{llrr}
\toprule
\textbf{Abbrev.} & \textbf{Category}& \llap{\textbf{ATL \%}}& \textbf{CLBL \%}\\
\midrule
ADUL & Other Adult Materials                &  0.91    &   0.77 \\
ARM  & Armed Forces                         &  0.76    &   0.31 \\ 
BLOG & Personal websites and blogs          &  2.00    &   8.97 \\
DOM  & Domain Parking                       &  0.21    &   0.28 \\ 
ENT  & Entertainment                        &  2.66    &   2.25 \\
FILE & File Sharing and Storage             &  1.89    &   0.55 \\
GAME & Games                                &  2.62    &   0.83 \\
GAMB & Gambling                             &  1.73    &   1.18 \\
HEAL & Health and Wellness                  &  2.02    &   1.04 \\
ILL  & Illegal or Unethical                 &  1.85    &   0.40 \\
IM   & Instant Messaging                    &  0.49    &   0.14 \\
IT   & Information Technology               &  9.31    &   4.17 \\
ITRA & Internet radio and TV                &  0.39    &   0.59 \\
LING & Lingerie and Swimsuit                &  0.76    &   0.14 \\
NEWS & News and Media                       &  10.03    &  18.87 \\
ORG  & General Organizations                &  6.82    &   4.77 \\
POL  & Political Organizations              &  1.56    &   5.28 \\
PORN & Pornography                          &  3.87    &   2.45 \\
PROX & Proxy Avoidance                      &  1.71    &   0.57 \\
RELI & Global Religion                      &  3.19    &   2.58 \\
SEAR & Search Engines and Portals           &  3.93    &   2.36 \\
SHOP & Shopping                             &  4.86    &   1.40 \\
SOC  & Social Networking                    &  1.19    &   1.34 \\
SOLI & Society and Lifestyles               &  0.76    &   0.97 \\
STRM & Streaming Media and Download         &  1.83    &   1.42 \\
\bottomrule
\end{tabular}
\end{table}

\subsection{Synthetic Analysis}\label{sec:synthetic}

Combining observations of all three types of censorship gives us a
clearer picture of what is censored in the countries we monitor,
and complements missing events in each.

We use the “FortiGuard” URL classification service, operated by
FortiNet~\cite{fortinet}, to categorize the contents of each test list.  This
service is sold as part of a “web filter” for corporations, which is the same
software as a nation-state censorship system, but on a smaller scale.  The
URLs on all our lists, together, fall into 79 high-level categories according
to this service; the 25 most common of these, for URLs that are censored at
least once, are listed in Table~\ref{tab:fortiguard_categories}, along with
the abbreviated names used in other tables in this section.

Table~\ref{category_testlist} shows the three most censored categories
of URLs for the five countries conducting the most censorship, based on
the percentage of unique URLs censored over time. It is divided into
four columns, showing how the results vary depending on which of our
test lists are considered: all of them, only ATL, only \mbox{CLBL-G}, or only
\mbox{CLBL-C}. Table~\ref{tab:category_table_full} in
Appendix~\ref{sec:detailed-results} continues this table with
information about the countries ranked 6 through 15.

Iran takes first place in all four columns, and Saudi Arabia is always
within the top three.  The other countries appearing in
Table~\ref{category_testlist} are within the top five only for some
test lists.  The top three categories blocked by each country change
somewhat from list to list.  For instance, pornography is much less
prominent on the country-specific lists than on the global
list. Iran's censorship is more uniformly distributed over topics than
the other countries, where censorship is concentrated on one or two
categories. These results demonstrate how the choice of test lists can
change observations about censorship policy.

\begin{figure}
\includegraphics[width=\hsize]{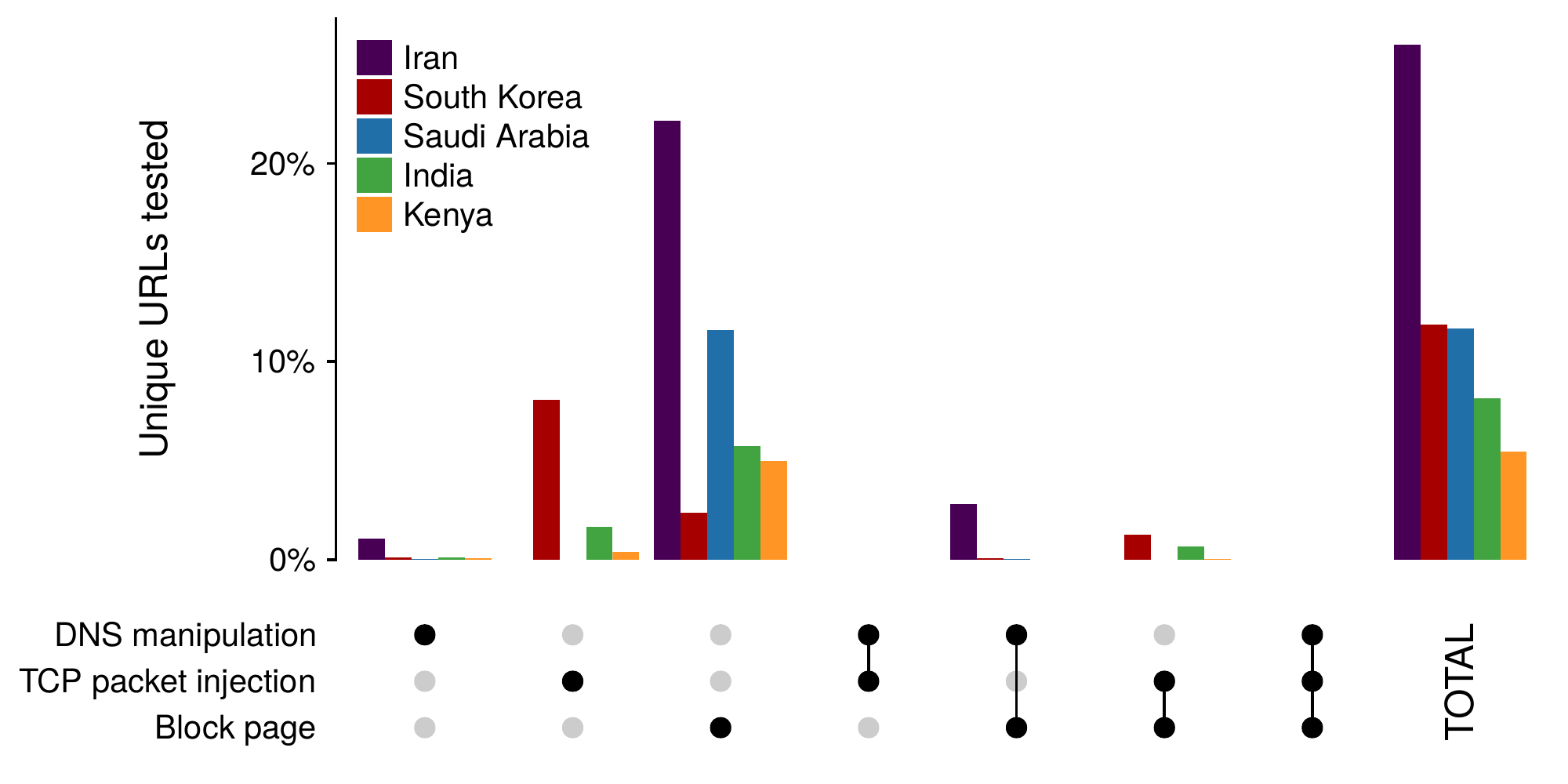}
\caption{\textsc{Combinations of Censorship Techniques}.  For the five
countries performing the most censorship overall, which combinations of the
three phenomena \systemname\ can detect are observed.  Except for “TOTAL,”
each group of bars is mutually exclusive---URLs counted under “DNS manipulation
and packet injection” are not also counted under either “DNS manipulation” or
“packet injection.”}
\label{fig:venn}
\end{figure}

Table~\ref{tab:technique_country} shows the top five countries
conducting the most censorship, for each of the three censorship
techniques that \systemname\ can detect, with the top three categories
censored with that technique.  This shows how censors use different
techniques to censor different types of content, as we mentioned in
\sectionref{sec:introduction}. For example, Turkey uses DNS
manipulation for categories ILL and STRM, but uses block pages for
PORN and NEWS.

Figure~\ref{fig:venn} shows how often the various blocking techniques
are combined. For instance, in Iran we detect some URLs being
redirected to a block page via DNS manipulation (comparing with
Table~\ref{tab:technique_country}, we see that these are the URLs in
the PORN and BLOG categories), but for many others, we detect only the
block page.  This could be because Iran uses a technique we cannot
detect for those URLs (\eg~route manipulation), or because our
analysis of packet injection is too conservative (see
Section~\ref{sec:packetinjection}).

\subsection{Longitudinal Analysis}\label{sec:longitudinal}

\begin{figure}
\includegraphics{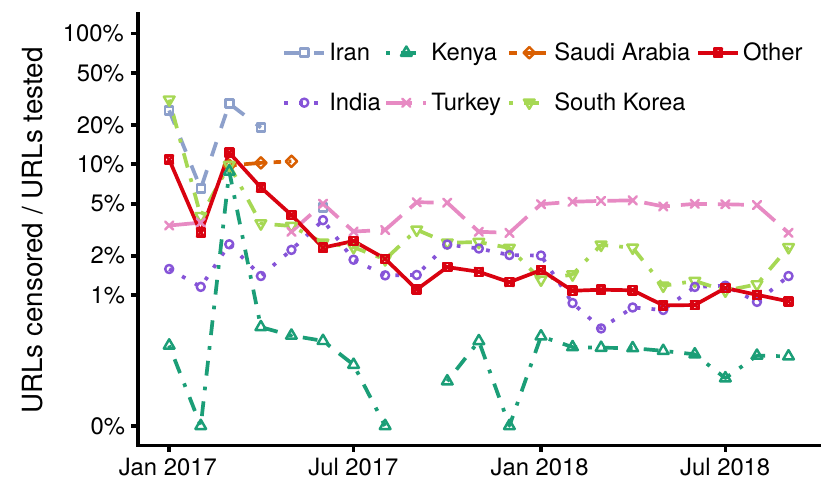}
\caption{\textsc{Logarithmic Plot of Longitudinal Trends}. Changes
  over time in the level of censorship, within the six countries where
  we observe the most censorship of URLs from ATL and \mbox{CLBL-G}, plus
  the aggregate of all other monitored countries.}
\label{fig:longitudinal_log}
\end{figure}

Collecting data for nearly two years gives us the ability to observe
changes in censorship over time. Figure~\ref{fig:longitudinal_log}
shows censorship trends for the six countries \systemname\ can monitor
that block the most URLs from the global test lists (ATL and \mbox{CLBL-G}),
plus a global trend line computed from aggregate measurements from all
the other monitored countries.  We do not have complete coverage for
Iran and Saudi Arabia, due to the outages mentioned
in~\sectionref{subsec:data_collection}.  The large dip in several of
the trend lines in February~2017 is an artifact due to month-to-month
churn within the Alexa rankings (see~\citet{Scheitle:2018:TopList}).

The global trend line shows a steady decreasing trend, which we
attribute to the rising use of secure channel protocols (\eg~TLS) and
circumvention tools.  This trend is also visible for South Korea but
not for the other top five countries.

Iran blocks 20--30\% of the URLs from ATL, more than any other
country.  This is due to extensive blocking of URLs in the NEWS and
BLOG categories.  Saudi Arabia consistently blocks roughly 10\% of ATL
and \mbox{CLBL-G} URLs, mostly from the PORN and GAMB categories with some
NEWS as well.  South Korea applies a similar level of blocking for the
PORN and GAMB categories; it is a much more democratic nation than
Saudia Arabia, but it nonetheless has applied draconian restrictions
to ``indecent Internet sites'' (including both pornography and
gambling sites) since before 2008~\cite{oni-kr}.

Censorship in Kenya is stable at a rate of roughly 0.4\% except for
March 2017, where the rate spikes to 10\%.  This is an artifact; for
that one month, our VOD in Kenya was connected to a network that
applied much more aggressive ``filtering'' to porn, gambling, and
proxy sites than is typical for Kenya, using a commercial product.

At the beginning of 2018, we observe a drop in the level of filtering
in India, from 2\% to 0.8\%, followed by a slow rise back to about
1.5\% after about four months. This coincides with political events:
India's telecommunications regulator announced support for ``net
neutrality'' at the end of 2017~\cite{India-telecom,
  Yadav2018:ooni-flaw}, and most ISPs suspended their filtering in
response. However, when a detailed regulation on net neutrality was
published in mid-2018~\cite{india.net.neutrality}, it became clear
that the government had not intended to relax its policy regarding
content deemed to be illegal, and the filtering was partially
reinstated.

Similarly, we see a rise in the level of filtering in Turkey in June
2017, from an earlier level of 3\% to 5\%.  Although it is not visible
on this chart, the topics censored also change at this time.  Prior to
the rise, most of the blocked sites in Turkey carried pornography and
other sexual content; after the rise, many more news sites were
blocked.  This, too, coincides with political events.  Following a
controversial referendum which increased the power of the Turkish
Presidency, the government has attempted to suppress both internal
political opposition and news published from other countries.
International news organizations took notice of the increased level of
Turkish online censorship in May of 2017~\cite{freedom-turkey,
  Turkey-wikipedia}, while \systemname\ detected it around the end of
April.

\subsection{Heuristic False Positives}\label{sec:falseposneg}

We manually reviewed the results of all of our heuristic detectors for
errors.  Manual review cannot detect false negatives, because we have
no way of knowing that we \emph{should} have detected a site as
censored, but false positives are usually obvious.  Here we discuss
the most significant cases we found, and how we adjusted the
heuristics to compensate.

\myparab{DNS Manipulation.}
We manually verified the detection results identified by each
heuristic.  The only heuristic producing false positives was the rule
for when a vantage point and control nodes receive addresses in
different ASes. As we mentioned in Section~\ref{sec:dnsmanipulation},
this heuristic gives a false positive rate on the order of $10^{-4}$
with the value of $\theta$ we selected.

\myparab{Packet injection.}
As with DNS manipulation, we manually reviewed the results of each
heuristic for false positives.  We found many false positives for RST
or ICMP unreachable instead of SYN-ACK, leading us to reclassify these
as only ``probable'' censorship and not include them in the synthetic
analysis above.  We also found cases in all of the categories where a
packet anomaly was only observed once, for a URL that seemed unlikely
to be censored from that vantage (\eg~connection disrupted to an
airline website from a VPN vantage in the USA).  We therefore discount
all cases where a packet anomaly has only been observed once for that
URL in that country.

\myparab{Block pages.}
Our set of regular expressions did initially produce some false
positives, for instance on news reportage of censorship, quoting the
text of a block page.  We manually reviewed all of the matches and
adjusted the regular expressions until no false positives remained.
It was always possible to do this without losing any true positives.

\section{Other Cases of Network Interference}\label{sec:casestudy}

In this section, we describe three cases of network interference
discovered with \systemname, that are different from the form of
censorship we set out to detect: Geoblocking by content
providers~(\sectionref{Sec:Geoblocking}), injection of a script to
fingerprint clients~(\sectionref{Sec:Korea}), and injected
malware~(\sectionref{Sec:Coinhive}).

\subsection{Geoblocking and HTTP 451}\label{Sec:Geoblocking}

HTTP status code~451, “Unavailable for Legal Reasons,” was defined in
2016 for web servers to use when they cannot provide content due to a
legal obstacle (\eg\ the Google restricts access to clients from Iran
to enforce US sanctions~\cite{McDonald:2018:403}) or requests from
foreign governments~\cite{rfc7725}.

We observe 23 unique websites that return status~451, from vantages in
21 countries.  Six of these cases appear to be
\nolinkurl{wordpress.com} complying with requests from Turkey and
Russia (for blogs related to political and religious advocacy).  Along
with the HTTP 451 status, they also serve a block page, explaining
that \nolinkurl{wordpress.com} is complying with local laws and court
orders.  Two more websites (both pornographic) were observed to return
status~451 from Russia, with HTTP server headers indicating the error
originates from the Cloudflare CDN, but without any explanation.
Since the adoption of the GDPR~\cite{regulation2016regulation} we have
observed a few sites returning status~451 when visited from European
countries.

Since status~451 is relatively new, the older, more generic status~403
(“Forbidden”) is also used to indicate geoblocking for legal reasons.
Applying the tag frequency clustering technique described in
Section~\ref{sec:detection:tagfreq} to the accompanying HTML, we were
able to discover six more URLs, in four countries, where status~403 is
used with a block page stating that access is prohibited from the
client's location.  Three of these were gambling sites, with the text
of the block page stating that the sites are complying with local
regulations.

We also observe a related phenomenon at the DNS level.  From a single
VPN server located in the USA, we observed \nolinkurl{netflix.com}
resolving to an IANA-reserved IP address, \texttt{198.18.0.3}.  This
could be Netflix refusing to provide their service to users behind a
VPN.

\subsection{User Tracking Injection}\label{Sec:Korea}

Our detector for block pages~(\sectionref{sec:blockpages}) flagged a
cluster of TCP payloads observed only in South Korea. Upon manual
inspection, these pages contained a script that would fingerprint the
client and then load the originally intended page.  We observed
injections of this script over a five-month period from Oct.\ 2016
through Feb.\ 2017, from vantage points within three major Korean
ISPs, into 5--30\% of all our test page loads, with no correlation
with the content of the affected page.  By contrast, censorship in
South Korea affects less than 1\% of our tests and is focused on
pornography, illegal file sharing, and North Korean propaganda.

These scripts could be injected by the VPN service, the ISPs, or one
or more of their transit providers.  The phenomenon resembles
techniques used by ad networks for recording profiles of individual
web users~\cite{acar2014web}.  This demonstrates the importance of
manual checking for false positives in censorship detection.  All of
the detection heuristics described in
Sections~\ref{sec:packetinjection} and~\ref{sec:blockpages} triggered
on these scripts, but they are not censorship.

\subsection{Cryptocurrency Mining Injection}
\label{Sec:Coinhive}

Our block page detector also flagged a set of suspicious responses
observed only in Brazil.  The originally intended page would load, but
it would contain malware causing the web browser to mine
cryptocurrency.  (As of mid-2018, this is a popular way to earn money
with malware~\cite{Konoth:2018:MIL}.)  We were able to identify the
malware as originating with a botnet infecting MikroTik routers
(exploiting CVE-2018-14847), initially seen only in
Brazil~\cite{Coinhive-spiderLabs} but now reported to affect more than
200,000 routers worldwide~\cite{Coinhive-ccn}.  Infected routers
inject the mining malware into HTTP responses passing through them.

The malware appears in \systemname's records as early as July 21st,
2018---ten days before the earliest public report on the MikroTik
botnet that we know of.  If \systemname's continuous monitoring were
coupled with continuous analysis and alerting (which is planned) it
could also have detected this botnet prior to the public report.  This
highlights the importance of continuously monitoring network
interference in general.

\section{Comparison with other Platforms}
\label{sec:analysis:comparison}

\begin{table*}[bt]
\caption{High-level comparison of \systemname\ with five other
    censorship monitoring platforms.}\label{tab:compare_platforms}%
\def\ORfootnotetext{Open Relays: Internet hosts that will relay a
censorship probe from researchers' computers without any prior arrangement.}%
\def\encorefootnotetext{Due to privacy concerns, Encore does not record
this information.}%
\def\oonifootnotetext{The OONI client can optionally collect packet
  traces.  However, OONI's servers do not record traces, due to
  privacy concerns.}%
\def\yes{$\checkmark$}%
\def\maybe{\textbigcircle}%
\def\no{\hbox{}}%
\centering\footnotesize
\begin{minipage}{496pt} 
\centering
\begin{tabular}{lcccccccrrr}
\toprule
\textbf{Platform} & \textbf{Packet}   & \multicolumn{3}{c}{\textbf{Vantage Point Types}} & \multicolumn{3}{c}{\textbf{Detection Capabilities}}            & \multicolumn{3}{c}{\textbf{Coverage (avg/max)}} \\
                  & \textbf{Capture}  & \textbf{VPNs} & \textbf{ORs}\footnote{\ORfootnotetext}  & \textbf{VODs} & \textbf{DNS} & \textbf{TCP} & \textbf{Blockpage} &\textbf{Countries} &\textbf{ASes} &\textbf{URLs} \\
\midrule
\rowcolor{gray}Encore~\cite{burnett2015.encore}            & \no  & \no  & \yes & \no  & \no  & \no  & \no  & Unknown\rlap{\footnote{\encorefootnotetext}}
                                                                                                                        & Unknown              & 23  \\
Satellite-Iris~\cite{Pearce:2017:Iris,Will:2016:Satellite} & \no  & \no  & \yes & \no  & \yes & \no  & \no  & 174 / 179 & 3,261 / 3,617        & 2,094 / 2,423   \\
\rowcolor{gray}Quack~\cite{VanderSloot:2018:Quack}         & \no  & \no  & \yes & \no  & \no  & \no  & \no  & 75 / 76   & 3,528 / 4,135        & 2,157 / 2,484   \\
OONI~\cite{ooni-paper}                                     & \maybe\rlap{\kern0.5pt\footnote{\oonifootnotetext}}
                                                                  & \no  & \no  & \yes & \yes & \no & \yes  & 113 / 156 & 670 / 2,015          & 13,582 / 20,258 \\
\rowcolor{gray}\systemname                                 & \yes & \yes & \no  & \yes & \yes & \yes & \yes & 42 / 50   & 48 / \iclabcountries & 16.964 / 23,992 \\
\bottomrule
\end{tabular}
\end{minipage}
\end{table*}

Other censorship measurement platforms active, at the time of writing, include
Encore~\cite{burnett2015.encore},
Satellite-Iris~\cite{Will:2016:Satellite, Pearce:2017:Iris},
Quack~\cite{VanderSloot:2018:Quack}, and OONI~\cite{ooni-paper}.
Table~\ref{tab:compare_platforms} shows the high-level features
provided by each of these platforms, and a comparison of their
country, AS, and URL coverage for the two-month period of August and
September 2018.  (August 1, 2018 is the earliest date for which data
from Quack and Satellite-Iris has been published.)  All platforms
suffer some variation from day to day in coverage, so we report both a
weekly average and the maximum number of covered countries, ASes, and
URLs.
%
While many of the platforms described in Table~\ref{tab:compare_platforms}
have chosen to emphasize breadth of country and AS coverage at the
expense of detail.  \Systemname\ takes the opposite approach,
collecting detailed information from a smaller number of vantage points.

\subsection{Quack}
\label{sec:Quack}

Quack relies on public echo servers to measure censorship. 
 It requires at least 15 echo servers within the same
country for robust measurements~\cite{VanderSloot:2018:Quack}.
Currently, these are available to Quack from 75 countries.  95 more
countries have at least one echo server, which can still provide
some measurements.

Quack aims to detect censorship of websites, but it does not
send or receive well-formed HTTP messages.  Instead it sends packets
that mimic HTTP requests, which the echo server will reflect back to
the client.  It expects the censor to react to this reflection in
the same way that it would to a real HTTP message.  The designers of
Quack acknowledge the possibility of false negatives when the censor
only looks for HTTP traffic on the usual ports (80 and 443).  More
seriously, manual inspection of the Quack data set reveals that in
32.6\% of the tests marked as \textit{blocked}, the client did not
successfully transmit a mimic request in the first place.  We have
reported this apparent bug to the Quack team.

\subsection{Satellite-Iris}
\label{sec:Satellite-Iris}

Satellite-Iris~\cite{censoredplanet} combines
Satellite~\cite{Will:2016:Satellite} and
Iris~\cite{Pearce:2017:Iris}. It focuses on DNS manipulation,
measuring from open DNS resolvers.  It compares the responses received
from these vantage points with responses observed from a control node.
It also retrieves corresponding TLS certificates from the
Censys~\cite{censys2015search} data set and checks whether they are
valid.  It applies several heuristics to each response, \emph{all} of
which must be satisfied for the response to be judged as censorship.
We now highlight two cases where their heuristics lead to false
negatives.

If Satellite-Iris can retrieve a TLS certificate from \emph{any} of
the IP addresses in the open resolver's response, and that certificate
is valid for \emph{any} domain name, it considers the response not to
be censored.  This means Satellite-Iris will not detect any case where
the censor supplies the address of a server for a different domain in
forged DNS responses.  In Satellite-Iris's published data set, 0.01\%
of the measurements are affected by this.  Specifically, measurements
from China and Turkey show domains belonging to the BBC, Google, Tor
and others resolving to an IP address belonging to Facebook's server
pool.

Despite the design bias toward false negatives, we also find that
83.8\% of the DNS responses considered censored by Satellite-Iris may
be false positives.  Satellite-Iris depends on the Censys data set to
distinguish DNS poisoning from normal IP variation
(\eg\ due to geotargeting and load balancing by CDNs).  When Censys
information is unavailable, it falls back on a comparison to a single
control resolver, which is inadequate to rule out normal variation, as
discussed in \sectionref{sec:dnsmanipulation}.

\subsection{OONI}
\label{sec:OONI}

\begin{figure}
\centering
\includegraphics[width=0.35\textwidth]{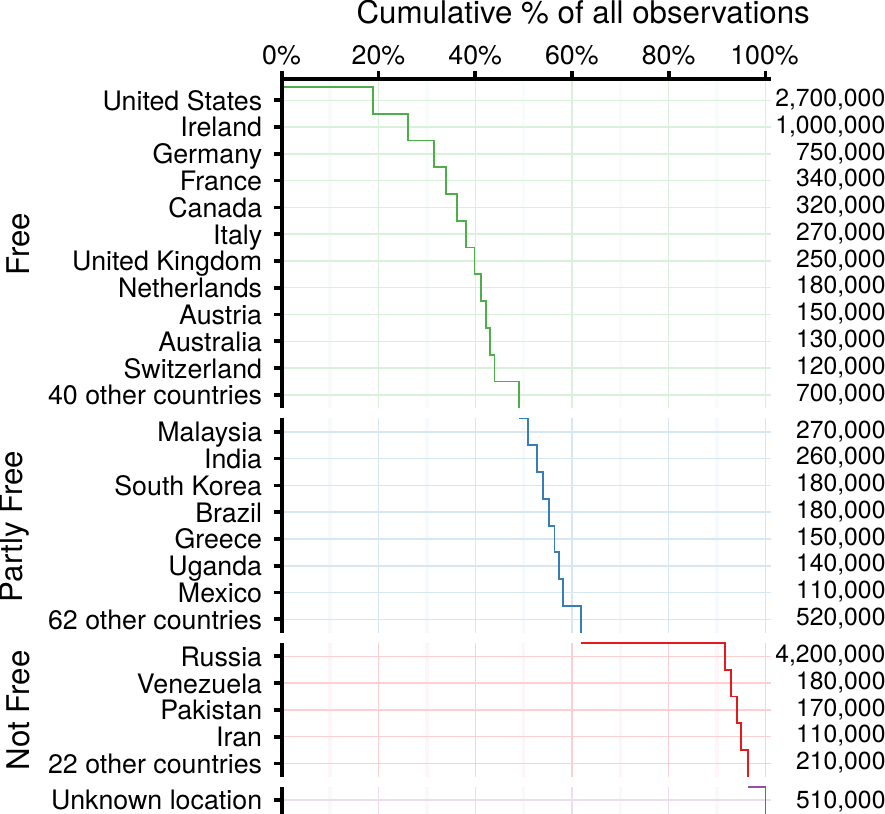}
\caption{Distribution of OONI observations by country in August and
  September of 2018, grouped by Freedom House classification.}
\label{fig:ooni_top_countries}
\end{figure}

OONI~\cite{ooni-paper} relies on volunteers who run a testing
application manually.  The application is available on all major
desktop and mobile operating systems except Windows.  In August and
September of 2018, OONI's volunteers conducted 14,000,000 measurements
from 156 countries, and reported 29,982 unique URL-country pairs as
blocked.

OONI's reliance on volunteers, and on manual operation of the testing
application, means that its coverage is not evenly distributed over
websites or countries.  Their “primary web connectivity” test suite
also tests ATL and \mbox{CLBL-G}, but the mobile phone version of the
application only tests a short subsample on each run, in order to
limit the time and bandwidth consumed by the test.  62\% of the
measurements for the August and September 2018 period tested fewer
than 80 URLs. Figure~\ref{fig:ooni_top_countries} shows the
distribution of countries covered by OONI. 86\% of all observations
originate from the 23 countries named in this figure, and 48\% are
from just two countries---Russia and the USA.  For another 4\% of the
observations, no location could be identified.  Comparing
Figure~\ref{fig:ooni_top_countries} to Figure~\ref{fig:scatter} shows
that OONI does not achieve better coverage of “partly free” and “not
free” countries than \systemname\ does.  Volunteers whose Internet
access is unreliable or misconfigured may submit inaccurate results.
Indeed, more than 100,000 of the observations are tagged inconclusive
due to local network errors (\eg~disconnection during the test).

Because OONI's testing application runs without any special
privileges, it normally cannot record packet traces.  OONI's detection
heuristics are rudimentary, leading to a high level of false
positives.  OONI's DNS consistency test will flag \emph{any
  disagreement} between the client's local recursive resolver and a
public DNS utility as censorship~\cite{ooni-dns}. OONI's block page
detector relies on the “30\% shorter than uncensored page” heuristic
proposed by~\citet{jones.2014.blockpages}, but innocuous server errors
are also short compared to normal page.

\citet{Yadav2018:ooni-flaw} reported very high levels of inaccuracy in
OONI's results for India, with an 80\% false positive rate and a
11.6\% false negative rate.  We confirm a high false positive rate
for OONI's block page detector: of the 12,506 unique anomalous HTTP
responses reported by OONI's volunteers in August and September 2018,
our block-page detector only classifies 3,201 of them as censorship,
for a 74.4\% false positive rate.  The most common cause of false
positives is a response with an empty HTML body, which can occur for a
wide variety of innocuous reasons as well as
censorship~\cite{McDonald:2018:403, afroz2018exploring,
  tschantz2018bestiary}.

\section{Related Work}\label{sec:related}

China's censorship practices have been studied in
detail~\cite{claytonchina, parkchina, xuchina, ensafi2015analyzing,
  wright2014cn.regional} due to the worldwide reputation of the “Great
Firewall” and the relative ease of gaining access to vantage points
within the country.  Other countries receiving case studies include
Iran~\cite{Aryan:2013, anderson2013ir.throttling},
India~\cite{Yadav2018:ooni-flaw}, Pakistan~\cite{pakistancensorship,
  khattak.2014.isp}, Syria~\cite{syriacensorship}, and Egypt and
Libya~\cite{dainotti2013eg.ly.outages}.  Whenever researchers have had
access to more than one vantage point within a country, they have
found that the policy is not consistently enforced.  There is always
region-to-region and ISP-to-ISP variation.

Broader studies divide into two lines of research.  One group of studies
investigate worldwide variation in censorship: for instance,
whether censorship mainly interferes with DNS lookups~\cite{Pearce:2017:Iris}
or subsequent TCP connections, and whether the end-user is informed of
censorship~\cite{verkamp2012.mechanics, hellmeier.2016.toolkit}. In some
cases, it has been possible to identify the specific software in
use~\cite{dalek2013.url.filtering, jones.2014.blockpages}. Another line of
work aims to understand what is censored and why~\cite{aase2012whiskey,
  burnett.2013.sense, weinberg.2017.topics}, how that changes over
time~\cite{anderson2014.ripe, gill.2015.worldwide}, how people react to
censorship~\cite{knockel.2011.skype, wright2016.filterprints}, and how the
censor might react to being monitored~\cite{burnett.2013.sense}.

We described the difficulties with relying on volunteers
in~\sectionref{sec:introduction} and~\sectionref{sec:iclab}.  Several groups
of researchers have sought alternatives.
CensMon~\cite{sfakianakis2011.censmon} used Planet~Lab nodes,
\citet{anderson2014.ripe} used RIPE~Atlas nodes, \citet{Pearce:2017:Iris} use
open DNS resolvers and \citet{VanderSloot:2018:Quack} use open echo servers.
\Citet{darer2017filteredweb} took advantage of the fact that the Chinese Great Firewall
will inject forged replies to hosts located outside the country.
 \Citet{burnett2015.encore}, \citet{ensafi2014cn.largescale}, and
\citet{pearce.2017.augur} all propose variations on the theme of using
existing hosts as reflectors for censorship probes, without the knowledge of
their operators, at different levels of the protocol stack.

Only a few studies have lasted more than a month.  Five prominent
exceptions are Encore \cite{burnett2015.encore}, IRIS
\cite{Pearce:2017:Iris}, OONI \cite{ooni-paper},
Quack\cite{VanderSloot:2018:Quack}, and Satellite
\cite{Will:2016:Satellite}, all of which share goals similar to
\systemname.  Section~\ref{sec:analysis:comparison} provides a
detailed comparison between \systemname\ and these projects.
Herdict~\cite{berkman.herdict} has also been active for years, but
simply aggregates user reports of inaccessible websites.  It does not
test or report why the sites are inaccessible.

\section{Limitations}\label{sec:discussion}

In this section, we discuss \systemname's limitations and how we have
addressed them.

\myparab{Discrimination against VPN users.} Some websites may block
access from VPN users~\cite{McDonald:2018:403,
  Will:2016:Satellite}. We sometimes observe this discrimination
against our VPN clients (see \sectionref{Sec:Geoblocking} for an
example), and are careful not to confuse it for censorship.

\myparab{Malicious VPN Providers.} Some VPN providers engage in
surveillance and traffic manipulation, for instance to monetize their
service by injecting advertisements into users'
traffic~\cite{Khan:2018:VPN}.  We avoid using VPN providers that are
known to do this.  Our block page detectors are designed not to
confuse dynamic content (\eg\ advertisements, localization) with
censorship, as described in~\sectionref{sec:blockpages}.  In
\sectionref{Sec:Korea} and~\ref{Sec:Coinhive} we describe surveillance
and malware injections that required manual inspection to distinguish
from censorship.

VPN providers are also known to falsely advertise the location of
their VPN servers~\cite{Weinberg:2018:CPL:Proxy}. We verify all server
locations using the technique described in
Appendix~\ref{sec:appendix_geoloc}.

\myparab{Bias in Test Lists.} ATL suffers from sampling bias and
churn~\cite{Scheitle:2018:TopList}.  \mbox{CLBL-G} and \mbox{CLBL-C} may suffer from
selection bias, since they are manually curated by activists.  We have
plans to revise the test lists and add more URLs as needed.

\mbox{CLBL-G} and \mbox{CLBL-C} are updated slowly.  It is not unusual
for more than half of the sites on a country-specific list to no
longer exist~\cite{weinberg.2017.topics}. This is not as much of a
limitation as it might seem, because censors also update their lists
slowly. Several previous studies found that long-gone websites may
still be blocked~\cite{aase2012whiskey, pakistancensorship,
  Yadav2018:ooni-flaw}.

\myparab{Coverage of “Not Free” Countries.} As discussed in
Section~\sectionref{sec:vantage.points}, the risks involved with
setting up many vantage points in certain sensitive (``not free'')
countries prevent us from claiming we can obtain complete coverage at
all times.  However, the set of countries \systemname\ covers gives us
a good, if imperfect, longitudinal overview of worldwide censorship.

\myparab{Evading Censorship Detection.}  Censors are known to try to
conceal some of their actions (``covert'' censorship).
\Systemname\ can detect some covert censorship, as discussed in
\sectionref{sec:detection}, but not all of it.  The “uncertain” and
“probably censored” cases of TCP packet injection
(Figure~\ref{f:packet-anomaly-matrix} in
\sectionref{sec:packetinjection}) are priorities for further
investigation.  Censors could further conceal their actions by
disabling filtering for IP addresses that appear to be testing for
censorship.  Comparing results for vantage points in the same country
gives us no reason to believe any country does this today.

\section{Conclusions}\label{sec:conclusion}

We presented \systemname, a global censorship measurement
platform that is able to measure a wide range of network interference and
Internet censorship techniques.

By using VPN-based vantage points, \systemname\ provides flexibility and
control over measurements, while reducing risks in measuring
Internet censorship at a global scale. Between January 2017 and September
2018, \systemname\ has conducted \iclabmeasurements\ measurements over
\iclaburls\ URLs in \iclabcountries\ countries.

\systemname\ is able to detect a variety of censorship mechanisms as
well as other forms of network interference. Other longitudinal
measurement platforms may have more vantage points and accumulated
data than \systemname, but also more errors, and/or they only focus on
a specific type of censorship.  Our platform can more reliably
distinguish normal network errors from covert censorship, and our
clustering techniques discovered 48 previously unknown block pages.

As we continue to operate \systemname\ and interact with relevant political
science and civil society organizations, \systemname\ will not
only make new technical observations, but also place qualitative work in this
area on a firm empirical footing.

\section*{Acknowledgements}

Nick Feamster,
Sathyanarayanan Gunasekaran,
Ben Jones,
Mose Karanja,
Anke Li,
the Berkman Klein Center,
the Small Media Foundation,
and the Citizen Lab, especially
Masashi Crete-Nishihata,
Jakub Dalek,
and Adam Senft,
have all provided invaluable assistance
with the implementation, testing, and deployment of \systemname.

We would like to thank our shepherd, Leyla Bilge,
and all of the anonymous reviewers for their feedback on this paper.
We also thank
Behtash Banihashemi,
Arun Dunna,
Pamela Griffith,
Steve Matsumoto,
Rishab Nithyanad,
Pinar Ozisik,
Vyas Sekar,
Mahmood Sharif,
Rachee Singh,
Kyle Soska,
Janos Szurdi,
Xiao Hui Tai,
and
Nicholas Weaver
for helpful comments and suggestions.

This research was financially supported
by the Ministry of Science and ICT, Korea, under award
IITP-2019-H8601-15-1011;
by the National Science Foundation, United States, under awards
CNS-1350720,
CNS-1651784,
CNS-1700657,
CNS-1740895,
and
CNS-1814817;
by a Google Faculty Research Award;
and by the Open Technology Fund under an Information Controls Fellowship.
The opinions in this paper are those of the authors and do not
necessarily reflect the opinions of the sponsors, nor of the
governments of the Republic of Korea or the United States of America.

\bibliography{bibliography}

\appendices

\section{Access to Different ASes within countries of interest}
\label{sec:appendix_networks_access}

\begin{figure}[bh!]
\includegraphics[width=8cm]{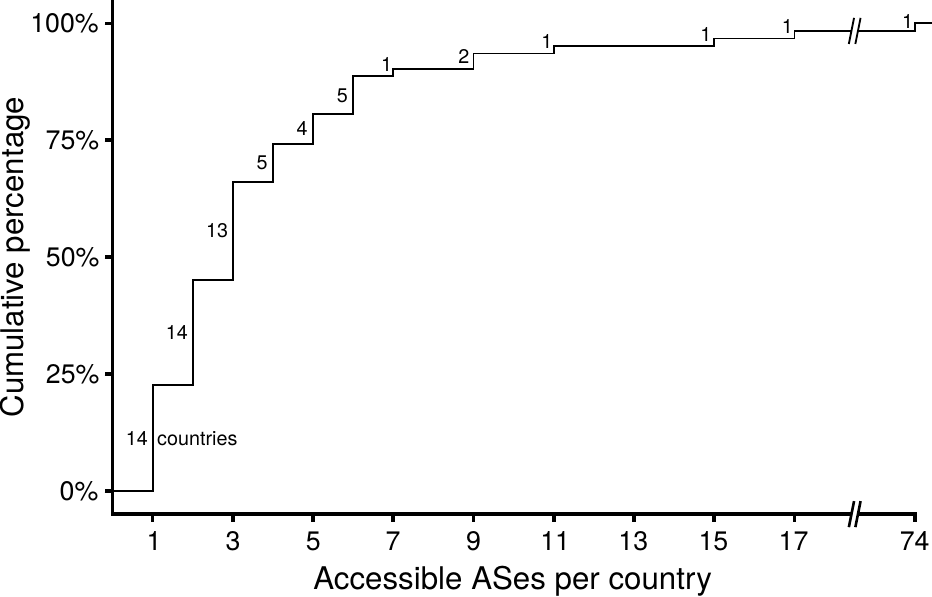}
\caption{CDF of number of accessible AS(es) per country}
\label{fig:asn_per_country}
\end{figure}

Censorship policies are known to vary from region to region and network to
network within a single country~\cite{ensafi2014cn.largescale,
  khattak.2014.isp, aase2012whiskey, wright2014cn.regional, xuchina}.
Therefore, comprehensive monitoring requires vantage points located in diverse
locations within a country.  Some VPN services offer servers in several
physical locations within a single country, making this simple.  Even when they
don't advertise several physical locations, we have found that they often
load-balance connections to a single hostname over IP addresses in several
different ASes and sometimes different physical locations as well.  When
possible, we increase diversity further by subscribing to multiple VPN
services.  Figure~\ref{fig:asn_per_country} shows a CDF of the number of
networks we can access in each country, combining all the above factors; we are
able to access two or more networks in 75\% of all countries, and three or more
networks in 50\%.

\section{VPN Proxy Location Validation}
\label{sec:appendix_geoloc}

Commercial VPN services cannot be relied on to locate all of their servers in
the countries where they are advertised to be~\cite{Weinberg:2018:CPL:Proxy}.
\systemname\ therefore checks the location of each VPN server before using it
for measurements.  We assume that packets are not able to travel faster than
153 km/ms ($0.5104\,c$) over long distances.  We measure the round-trip time
from each VPN server to a set of landmark hosts in known locations, drawn from
the RIPE Atlas measurement constellation~\cite{ripe.2015.atlas}.  If any packet
would have had to travel faster than 153 km/ms to reach the advertised country
and return in the measured time, we assume the server is not in its advertised
location, and we do not use it as a vantage point.

The VPN services we subscribe to collectively advertise endpoints in 216
countries.  Our checker is only able to confirm the advertised location for
\iclabvpnscountries\ countries (25.5\% of the total). Compared to the results
reported by \citet{Weinberg:2018:CPL:Proxy}, who tackled the same problem with
more sophisticated techniques, our method rejects significantly more proxies
(10\% more on average when we experiment across multiple providers).  Possibly
some of those proxies could be used after all, but we do not want to attribute
censorship to the wrong country by accident, so we are being cautious.

\section{Freedom House and Reporters Without Borders Scores}
\label{sec:appendix_fscore}

\begin{table*}
    \caption{Censorship by Test List and Category\ulcshape, continued.
    For each of the three types of test list we use (see
    Section~\ref{s:test_lists}), the next ten countries performing
    the most censorship of URLs from that country, the top three
    FortiGuard categories among
    their censored URLs (abbreviations defined in
    Table~\ref{tab:fortiguard_categories}), and the percentage of all
    censored URLs from that category.  We only observe 14 countries to
    censor anything from \mbox{CLBL-C}.}
    \label{tab:category_table_full}
    \label{category_testlist_full}
    \centering\footnotesize
    {\def\nsep{\hskip 0.75\tabcolsep}
    \begin{tabular}{ll@{\nsep}rll@{\nsep}rll@{\nsep}rll@{\nsep}r}
    \toprule
    \multicolumn{3}{c}{\textbf{Overall}} &
    \multicolumn{3}{c}{\textbf{Alexa Global (ATL)}} &
    \multicolumn{3}{c}{\textbf{Globally Sensitive (\mbox{CLBL-G})}} &
    \multicolumn{3}{c}{\textbf{Per-Country Sensitive (\mbox{CLBL-C})}} \\
    \textbf{Country} & \textbf{Category} & \textbf{Pct.} &
    \textbf{Country} & \textbf{Category} & \textbf{Pct.} &
    \textbf{Country} & \textbf{Category} & \textbf{Pct.} &
    \textbf{Country} & \textbf{Category} & \textbf{Pct.} \\
    \cmidrule(r){1-3}\cmidrule(r){4-6}\cmidrule(r){7-9}\cmidrule{10-12}
    Turkey          & PORN & 40.2\% & Kenya        & ILL  & 28.1\%   & India        & NEWS & 10.3\%  & South Korea & PORN  & 16.7\% \\
                    & GAMB & 16.6\% &              & PORN & 25.0\%   &              & ILL  &  9.2\%  &             & NEWS  & 16.1\% \\
                    & NEWS &  9.2\% &              & GAME  &  6.2\%  &              & IT   &  8.0\%  &             & SHOP  & 11.8\% \\
    \cmidrule(r){1-3}\cmidrule(r){4-6}\cmidrule(r){7-9}\cmidrule{10-12}
    Russia          & GAMB & 23.4\% & Russia       & PORN & 26.3\%   & United States& NEWS & 8.0\%   & China       & NEWS  & 46.1\% \\
                    & PORN & 10.0\% &              & SHOP & 21.0\%   &              & IT   & 6.9\%   &             & ORG   & 46.1\% \\
                    & NEWS &  7.6\% &              & STRM & 10.5\%   &              & SEAR & 6.3\%   &             & RELI  &  7.7\% \\
    \cmidrule(r){1-3}\cmidrule(r){4-6}\cmidrule(r){7-9}\cmidrule{10-12}
    Uganda          & PORN & 42.6\% & Japan        & SEAR & 19.0\%   & Uganda       & PORN  & 42.6\%  & Hong Kong  & ORG  & 100.0\% \\
                    & ADUL & 11.7\% &              & NEWS &  9.5\%   &              & ADUL  & 11.7\%  &            &      &   \\
                    & LING & 10.3\% &              & GAME &  9.5\%   &              & LING  & 10.3\%  &            &      &   \\
    \cmidrule(r){1-3}\cmidrule(r){4-6}\cmidrule(r){7-9}\cmidrule{10-12}
    Netherlands     & NEWS & 13.4\% & Netherlands  & ILL  &  15.3\%  & Russia       & GAMB  & 39.4\% & Poland      & GAMB  & 100.0\% \\
                    & ILL  &  8.5\% &              & NEWS &  15.3\%  &              & PORN  & 14.9\% &             &   &  \\
                    & SEAR &  8.5\% &              & SEAR &  15.3\%  &              & RELI  &  5.3\% &             &  &  \\
    \cmidrule(r){1-3}\cmidrule(r){4-6}\cmidrule(r){7-9}\cmidrule{10-12}
    Japan           & NEWS & 11.0\% & Sweden       & SEAR & 27.2\%   & Netherlands  & NEWS  & 13.0\% & Singapore   & PROX  & 66.6\% \\
                    & GAME &  9.6\% &              & BLOG &  9.1\%   &              & ILL   &  7.2\% &             & GAME  & 33.3\% \\
                    & SEAR &  9.6\% &              & STRM &  9.1\%   &              & GAME  &  7.2\% &             &   &   \\
    \cmidrule(r){1-3}\cmidrule(r){4-6}\cmidrule(r){7-9}\cmidrule{10-12}
    Australia       & SEAR & 15.4\% & Hong Kong    & STRM & 20.0\%   & Japan        & NEWS  & 11.5\% & Ukraine     & BLOG  & 75.0\% \\
                    & ILL  & 10.7\% &              & SEAR & 20.0\%   &              & GAME  &  9.8\% &             & NEWS  &  8.3\% \\
                    & NEWS &  9.2\% &              & SHOP & 10.0\%   &              & ILL   &  6.5\% &             & IT    &  8.3\% \\
    \cmidrule(r){1-3}\cmidrule(r){4-6}\cmidrule(r){7-9}\cmidrule{10-12}
    Sweden          & GAME & 10.3\% & Australia    & ILL  & 30.0\%   & Australia    & SEAR  & 14.5\% & Malaysia    & PORN  & 100.0\% \\
                    & NEWS & 10.3\% &              & SEAR & 20.0\%   &              & NEWS  & 10.9\% &             &   &  \\
                    & STRM &  6.9\% &              & SHOP & 10.0\%   &              & ILL   &  7.2\% &             &   &   \\
    \cmidrule(r){1-3}\cmidrule(r){4-6}\cmidrule(r){7-9}\cmidrule{10-12}
    New Zealand     & GAME & 11.5\% & New Zealand  & SEAR & 20.0\%   & Sweden       & GAME  & 10.6\% & Colombia    & ITRA  & 100.0\% \\
                    & HEAL &  9.6\% &              & GAME & 20.0\%   &              & ILL   &  6.4\% &             &   &  \\
                    & SEAR &  9.6\% &              & ILL  & 10.0\%   &              & STRM  &  6.4\% &             &   &   \\
    \cmidrule(r){1-3}\cmidrule(r){4-6}\cmidrule(r){7-9}\cmidrule{10-12}
    China           & NEWS & 17.0\% & United States& SEAR & 21.6\%   & Hong Kong    & NEWS  & 10.9\% & Brazil      & SOLI  & 100.0\% \\
                    & ORG  & 12.7\% &              & NEWS & 10.8\%   &              & GAME  & 10.8\% &             &   &  \\
                    & SEAR &  6.4\% &              & ILL  &  8.1\%   &              & STRM  &  8.7\% &             &   &   \\
    \cmidrule(r){1-3}\cmidrule(r){4-6}\cmidrule(r){7-9}\cmidrule{10-12}
    Bulgaria        & ILL  & 11.6\% & China        & SEAR & 33.3\%   & New Zealand  & HEAL  &  9.5\% &             &   &  \\
                    & HEAL &  9.3\% &              & GAME & 16.6\%   &              & GAME  &  9.5\% &             &   &  \\
                    & GAME &  9.3\% &              & HEAL & 16.6\%   &              & NEWS  &  7.1\% &             &   &   \\
    \bottomrule
    \end{tabular}}
    \end{table*}

\begin{figure*}[t!]
\centering
\begin{tabular}{p{3.2in}@{\quad}p{3.4in}}
\multicolumn{1}{c}{\bfseries HTML structure}&
\multicolumn{1}{c}{\bfseries Visible message}\\
\cmidrule(r){1-1}\cmidrule{2-2}
{\raggedright\footnotesize\ttfamily\noindent
ACK+PSH\\
HTTP/1.1 200 OK\\
Connection:\ close\\
Content-Length:\ \textit{\textcolor{blue}{nnnn}}\\
Content-Type:\ text/html; charset="utf-8"\\
<!DOCTYPE html PUBLIC "-//W3C//DTD HTML 4.01//EN">\\
<html>\\
<head<title></title></head>\\
<body><h0><font color="black">\\
\textit{\textcolor{blue}{visible message}}\\
</font></h0></body></html>\par}&
{\raggedright\footnotesize\ttfamily\noindent
“This URL has been blocked under instructions of a\\
competent Government Authority or in compliance with\\
the orders of a Court of competent jurisdiction.\\
\smallskip
***This URL has been blocked under Instructions of the\\
Competent Government Authority or Incompliance to\\
the orders of Hon'ble Court.*** {\normalfont\itshape [sic]}\\
\smallskip
*“Error 403: Access Denied/Forbidden”*\\
\smallskip
404. That's an error.\\
\smallskip
HTTP Error 404 - File or Directory not found\\
\smallskip
HTTP Error 404 - File or Directory not found = \textit{\textcolor{blue}{http://...}}\par}\\
\end{tabular}
\vspace{-3ex}
\caption[Example cluster of block pages]
{\textsc{Example cluster of block pages.}  All of the messages in the
right-hand column were observed with the HTTP response headers and HTML
structure shown on the left.}
\label{fig:bp-cluster}
\end{figure*}

The international organization Freedom House, which promotes civil liberty and
democracy worldwide, issues a yearly report on “freedom on the Net,” in which
they rate 65 countries on the degree to which online privacy and free exchange
of information online are upheld in that country~\cite{freedom}. Each country
receives both a numerical score and a three-way classification: 16 of the 65
countries are considered “free,” 28 are “partly free,” and 21 are “not free.”
Unfortunately, 33 of the countries studied by \systemname\ are not included in
this report.

The international organization Reporters Without Borders (RWB) issues a
similar yearly report on freedom of the press. This report covers
189~countries and territories, including all 65~of the countries rated by
Freedom House, and all \iclabcountries\ of the countries studied by
\systemname~\cite{rwb} Each country receives a numerical score and a color
code (best to worst: 16~countries are coded white, 42~yellow, 59~orange,
51~red, and 21~black). Press freedom is not the same as online freedom, and
the methodologies behind the two reports are quite different, but the scores
from the two reports are reasonably well correlated (Kendall's $\tau = 0.707$,
$p \approx 10^{-16}$). We used a simple linear regression to map RWB scores
onto the same scale as FH scores, allowing us to label all of the countries
studied by \systemname\ as “free” (72), “partly free” (85), or “not free” (32)
in the same sense used by Freedom House.

\section{Detailed censorship results}\label{sec:detailed-results}

Table~\ref{tab:category_table_full} continues
Table~\ref{tab:category_table} (Section~\ref{sec:analysis:results}),
showing countries 6 through 15 in the same ranking, with the top three
FortiGuard categories among their censored URLs, and the percentages of
all their censored URLs within those categories.

Many of these countries censor only a few of the URLs on the lists we
tested, so Table~\ref{tab:category_table_full} may reflect biases of
the test lists, such as over-representation of the GAME, IT, NEWS, and
SEAR categories on both ATL and \mbox{CLBL-G}.
\balance

The presence of countries such as Japan, the Netherlands, Sweden, and
the United States in this table could indicate that our detectors
still have false positives, or that individual ISPs and/or the VPN
services we are using are blocking access to certain sites.  The
latter is likely for the ILL category, which includes many sites that
facilitate copyright infringement.

Figure~\ref{fig:bp-cluster} shows an example group of block pages
detected by textual similarity clustering, including two variations on
the Indian legal jargon mentioned in
Section~\ref{sec:detection:clustering}, but also messages mimicking
generic HTTP server errors.  This demonstrates how similarity
clustering can detect covert as well as overt censorship.

\end{document}